\documentclass[final,3p,12pt,sort&compress]{elsarticle}
\usepackage[T1]{fontenc}
\usepackage[utf8]{inputenc}
\usepackage{url}

\usepackage{graphicx}
\usepackage{xcolor}
\usepackage{multirow}

\usepackage{amssymb,amsmath}

\usepackage{float}
\usepackage{caption}
\usepackage{subcaption}

\usepackage[strings]{underscore} 

\usepackage[normalem]{ulem}

\begin{document}

\begin{frontmatter}

\title{Using Space-Filling Curves and Fractals to Reveal Spatial and Temporal Patterns in Neuroimaging Data}

\author[1,2]{Jacek Grela\corref{cor1}}
\ead{jacek.grela@uj.edu.pl}
\author[1]{Zbigniew Drogosz\corref{cor1}}
\ead{zbigniew.drogosz@alumni.uj.edu.pl}
\author[1]{Jakub Janarek}
\author[1,2]{Jeremi K. Ochab\corref{cor1}}
\ead{jeremi.ochab@uj.edu.pl}
\author[3]{Ignacio Cifre} 
\author[1,2]{Ewa Gudowska-Nowak}
\author[1,2]{Maciej A. Nowak}
\author[1,2,4]{Paweł Oświęcimka\corref{cor1}}
\ead{pawel.oswiecimka@ifj.edu.pl}
\author[5,6]{Dante R. Chialvo}
\author[]{for the Alzheimer’s Disease Neuroimaging Initiative\corref{cor2}}
\cortext[cor1]{Corresponding author}
\cortext[cor2]{Data used in preparation of this article were obtained from the Alzheimer’s Disease Neuroimaging Initiative (ADNI) database (adni.loni.usc.edu). As such, the investigators within the ADNI contributed to the design and implementation of ADNI and/or provided data but did not participate in analysis or writing of this report. A complete listing of ADNI investigators can be found at: http://adni.loni.usc.edu/wp-content/uploads/how_to_apply/ADNI_Acknowledgement_List.pdf}

\address[1]{Institute of Theoretical Physics, Jagiellonian University, 30-348 Kraków, Poland}
\address[2]{Mark Kac Center for Complex Systems Research, Jagiellonian University, 30-348 Kraków, Poland}
\address[3]{Facultat de Psicologia, Ci\`encies de l'educaci\'o  i de l'Esport, Blanquerna, Universitat Ramon Llull,  	Barcelona, Spain.}
\address[4]{Complex Systems Theory Department, Institute of Nuclear Physics, Polish Academy of Sciences, 31-342 Kraków, Poland}
\address[5]{Center for Complex Systems \& Brain Sciences (CEMSC$^3$), Escuela de Ciencia y Tecnolog\'ia,	Universidad Nacional de San Mart\'{i}n, Buenos Aires, Argentina}
\address[6]{Consejo Nacional de Investigaciones Cient\'{i}ficas	y Tecnol\'{o}gicas (CONICET), Buenos Aires, Argentina}

\begin{abstract}
\textit{Objective}
Magnetic resonance imaging (MRI), functional MRI (fMRI) and other neuroimaging techniques are routinely used in medical diagnosis, cognitive neuroscience or recently in brain decoding. They produce three- or four-dimensional scans reflecting the geometry of brain tissue or activity, which is highly correlated temporally and spatially. While there exist numerous theoretically guided methods for analyzing correlations in one-dimensional data, they often cannot be readily generalized to the multidimensional geometrically embedded setting.

\textit{Approach}
We present a novel method, Fractal Space-Curve Analysis (FSCA), which combines Space-Filling Curve (SFC) mapping for dimensionality reduction with fractal Detrended Fluctuation Analysis (DFA). We conduct extensive feasibility studies on diverse, artificially generated data with known fractal characteristics: the fractional Brownian motion, Cantor sets, and Gaussian processes. We compare the suitability of dimensionality reduction via Hilbert SFC and a data-driven alternative. FSCA is then successfully applied to real-world MRI and fMRI scans.

\textit{Main results}
The method utilizing Hilbert curves is optimized for computational efficiency, proven robust against boundary effects typical in experimental data analysis, and resistant to data sub-sampling. It is able to correctly quantify and discern correlations in both stationary and dynamic two-dimensional images. In MRI Alzheimer's dataset, patients reveal a progression of the disease associated with a systematic decrease of the Hurst exponent. In fMRI recording of breath-holding task, the change in the exponent allows distinguishing different experimental phases.

\textit{Significance}
This study introduces a robust method for fractal characterization of spatial and temporal correlations in many types of multidimensional neuroimaging data. Very few assumptions allow it to be generalized to more dimensions than typical for neuroimaging and utilized in other scientific fields. The method can be particularly useful in analyzing fMRI experiments to compute markers of pathological conditions resulting from neurodegeneration. We also showcase its potential for providing insights into brain dynamics in task-related experiments.

\end{abstract}


\end{frontmatter}

\section{Introduction}

Technological advancements in experimental science have resulted in an abundance of multidimensional data. This trend is particularly observed in neuroscience, where brain imaging techniques such as functional magnetic resonance imaging (fMRI) and positron emission tomography (PET) produce four-dimensional scans (comprising three spatial dimensions and one temporal dimension). Such developments require novel, dedicated methods of analysis combining the diversity of input data with interpretable outcomes.

In brain imaging, most established data analysis methods are tailored to specific imaging techniques—for example, ICA is mainly applied to functional MRI. In contrast, our method offers a nonparametric, general approach for quantifying the fractal properties of scans. According to the structural MRI method classification discussed in \cite{sMRIoverview2018}, our method generates a texture-based, statistical, second-order biomarker. This data-agnostic method aims to provide features similar to ROI-based biomarkers, such as those based on gray-matter volume, cortical thickness, or Voxel-Based Morphometry.

One compelling strategy of multidimensional data analysis involves transforming the data into more typical forms, such as a one-dimensional array, to enable standard time series methods. A fitting family of transformations is the Space-Filling Curves (SFCs), an indispensable tool of contemporary scientific computing~\cite{bader2012spacefilling}. SFCs are continuous functions that can convert multidimensional geometrically embedded data grids into one-dimensional sequences while preserving the local properties of the multidimensional data. This locality, or clustering, property of SFCs is highly desirable, ensuring that nearby points in a multidimensional space correspond to nearby points in a one-dimensional domain. This property renders SFCs prevalent instruments for dimensionality reduction and image compression~\cite{Owczarek2019SFCDimReduction}, efficient storage and retrieval of data from geographical information systems, cryptography, and data visualization~\cite{Aluru2004,Sheidaeian2011,Murali2019}. In neuroscience, SFC-based techniques also find extensive use in the processing of magnetic resonance imaging (MRI) data~\cite{Seginer2023SFCBrain1, Sharma2020SFCBrain2, Sakoglu2020SFCBrain3, Kontos2003SFCBrain4, Zhou2021SFCDataDriven}, e.g., selecting brain's activation maps from schizophrenia or Alzheimer's study. 

Mapping geometrically embedded multidimensional data onto a one-dimensional sequence presents an opportunity for utilizing fractal analysis of time series. One aspect of the theory of fractals explores the self-similar properties of data quantified by the fractal dimension exponent. In recent years, fractal characteristics have emerged across various scientific and artistic domains, spanning physics~\cite{pietronero2012,subramaniam2008}, biology~\cite{Buldyrev2009}, neuroscience~\cite{tagliazucchi2012FPCriticality,franca2018,ochab2022,WATOREK2024105916}, chemistry~\cite{stanley1988}, economics~\cite{Grau2001,oswiecimka2005,gao2022}, quantitative linguistics~\cite{montemurro2002FLongrange,drozdz2016}, and music~\cite{jafari2007}. The ubiquity of fractal structures owes much to the advancement of robust algorithms, among which detrended fluctuation analysis (DFA) is especially prominent~\cite{peng1994}. DFA quantifies the temporal organization of data through the Hurst exponent, a scaling exponent closely related to the fractal dimension. Thus, as scaling exponents reflect the correlation organization of data, DFA output can be interpreted through an interdisciplinary lens of fractality, thereby broadening the scope for understanding findings within the context of scale-free dynamics.

\begin{figure}
    \centering
    \includegraphics[width=0.7\columnwidth]{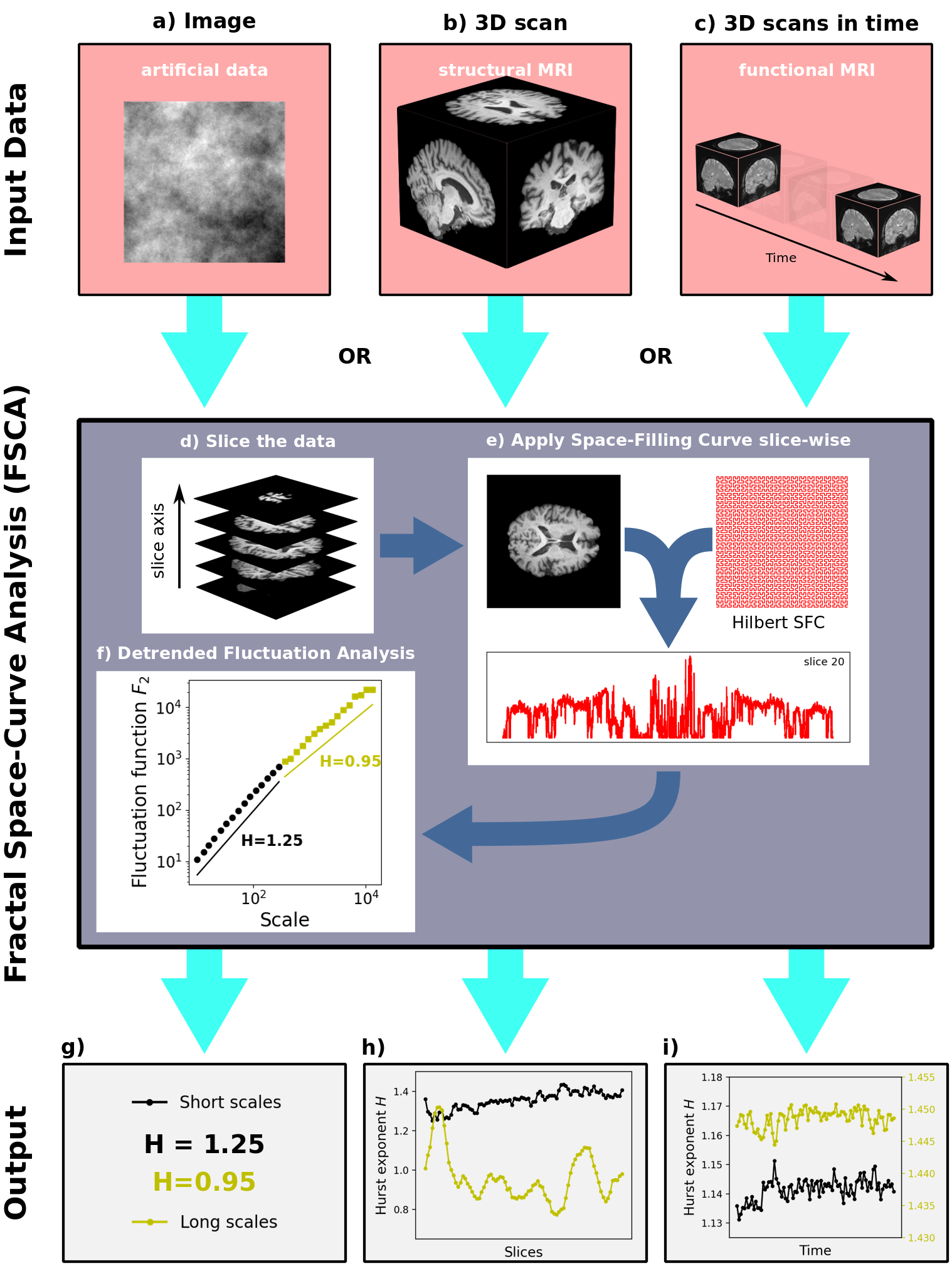}
    \caption{The flowchart of the proposed Fractal Space-Curve Analysis comprised of d) a slicing preprocessing step, e) space-filling-curve linearization mapping to reduce problem dimensionality and f) the detrended fluctuation analysis to extract Hurst exponents. The method is applicable to a) two-dimensional, b) three-dimensional, or c) three-plus-one-dimensional data. Method's output describes the fractality of the data in the form of g) single Hurst exponents, h) Hurst exponent profile resolved along a spatial axis or i) Hurst profile resolved in time.
    }
    \label{fig01:Flowchart}
\end{figure}

In this contribution, we introduce the Fractal Space-Curve Analysis (FSCA) shown in Fig.~\ref{fig01:Flowchart}, a novel approach to multidimensional data analysis integrating space-filling curves with detrended fluctuation analysis. FSCA utilizes SFCs as a method for locality-preserving dimensionality reduction, while DFA is employed to extract interpretable fractal properties from the data. In this work, we provide an extensive feasibility study of the method.
We find FSCA robust to the boundary effects, behaving consistently for different data resolutions and recreating correct fractal properties of synthetic data.    
We show its versatility in application to both the synthetic data and real-world data of diverse shapes spanning two-dimensional images, three-dimensional scans, and four-dimensional scans changing in time.  

The paper is structured as follows: Section~\ref{sec:Methods} introduces the concept of the proposed Fractal Space-Curve Analysis (FSCA) methodology. We begin by discussing the basic concepts of space-filling curves and fractal analysis using detrended fluctuation analysis. Section~\ref{sec:synthetic} presents an FSCA analysis of synthetic data that mimic the structure of neuroscientific datasets. Within this framework, fractal signals are considered to be two-dimensional fractional Brownian motion images, mirroring the analysis of MRI scans, binary Cantor sets, and two-plus-one-dimensional Gaussian processes similar to fMRI data analysis. We scrutinize FSCA for the optimal selection of the space-filling curve, potential boundary effects in experimental data, and data sampling density. In Section~\ref{sec:neuroimaging_data}, we apply FSCA to analyze neuroimaging data. We examine MRI datasets obtained from patients diagnosed with Alzheimer's disease, as well as fMRI signals from a single healthy individual undergoing various tests, referred to as the Poldrack dataset. Section~\ref{sec:conclusions} presents summary and conclusions.

\section{Method: Fractal Space-Curve Analysis (FSCA)}
\label{sec:Methods}

This chapter presents the main components of the proposed Fractal Space-Curve Analysis (FSCA). The stages of the method are illustrated by the flowchart in Fig.~\ref{fig01:Flowchart}. 
The data of any of several possible types—(a–c) in Fig.~\ref{fig01:Flowchart}—are first cut into two-dimensional slices (d), then each of the slices is turned into a time series with the help of the Hilbert space-filling curve (e), and, finally, detrended fluctuation analysis is applied to each time series to extract the Hurst exponents. The generated interpretable output is of a type depending on the dimensionality of the input data (g–i).

FSCA is easy to use, can be automatized and work ``out of the box'', and the knowledge of technical details is not required from the end user. In this section, however, we give a historical and mathematical description of space-filling curves (SFCs), together with their discrete versions, and a short introduction to DFA, a cornerstone of time-series fractal analysis.
In particular, we note the freedom of applying to FSCA space-filling curves other than the Hilbert SFC and the possible occurrence of disruptive boundary effects, which are typical in empirical data analysis and ought to be assessed through some approach. We also note that the SFCs used in data analysis stop at a finite number of iterations, which need not be the maximal value possible for a given data set, whence the notion of ``granularity''. These discussions will set the stage for some of the numerical tests in the next section (Sec.~\ref{sec:synthetic}), which—to anticipate—show that FSCA is robust, the Hilbert space-filling curve is recommended over other choices of space-filling curves that we tested, and the impact of various choices of granularity and boundary effects is low.

\subsection{Linearization via Space-Filling Curves}

A space-filling curve is a surjective continuous mapping from a one-dimensional interval onto a higher-dimensional region, typically exemplified by a mapping of the unit interval onto the unit square $[0, 1]\rightarrow[0, 1] \times [0,1]$ (see, e.g.,~\cite{Bader2011}). Although the cardinalities of these two sets are equal, Peano's discovery in 1890~\cite{peano1890surunecourbe} surprised the mathematical world as the first explicit example of such a mapping. This paradoxical object, Peano's space-filling curve, earned the name of a ``topological monster''. In each iteration, the fundamental shape (called the generator) is rescaled and appropriately repeated, while the curve itself is defined as the limit of this procedure as the number of iterations goes to infinity. In the following year, the second example of a space-filling curve was demonstrated by Hilbert~\cite{hilbert1891linie}, including also a picture of the first few iterations of the curve. Space-filling curves were long considered as a mere mathematical curiosity, but the rise of the theory of fractals, and especially the work of Mandelbrot~\cite{mandelbrot1968, mandelbrot1982fractal}, brought them into prominence, as they exemplified the fractal structures created by iterative hierarchical algorithms. Over time, a family of methods for constructing space-filling curves has been proposed, accompanied by a wide range of practical applications~\cite{bader2012spacefilling}. However, the possibility of combining them with fractal analysis for time series data remained largely unexplored.


\subsubsection{Hilbert and Peano curves}
\begin{figure}[H]
    \centering
    \includegraphics[width=1.0\columnwidth]{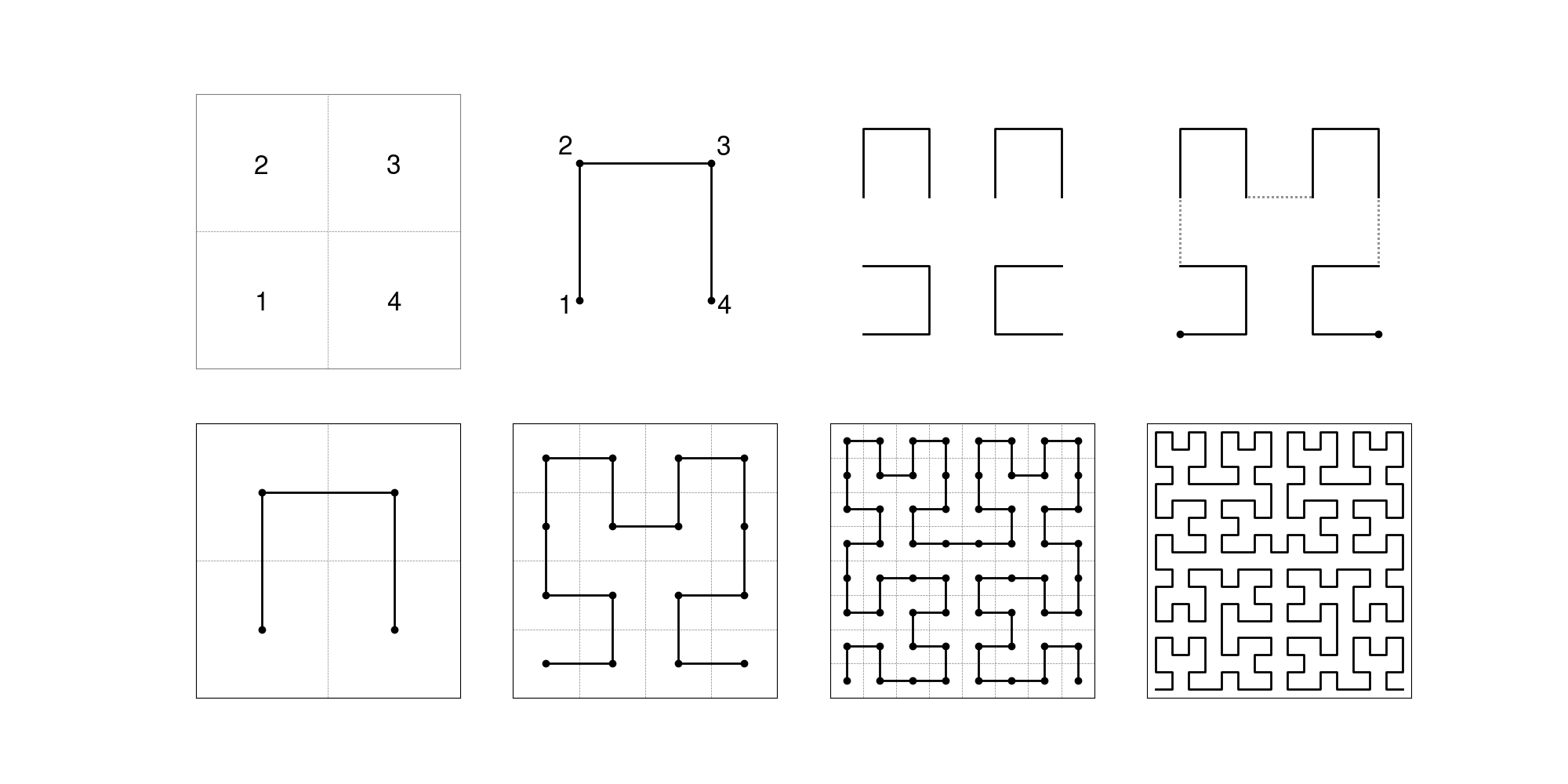}
    \caption{Iterative generation of Hilbert space-filling curves.}
    \label{fig02:HilbertPeanoExamples}
\end{figure}
The construction of the Hilbert curve is easiest to explain with the help of a picture (Fig.~\ref{fig02:HilbertPeanoExamples}). In the zeroth iteration, the unit interval is mapped into three contiguous line segments in the shape of the letter $\Pi$. The midpoint of the shape coincides with the center of the square. In the first iteration, the unit interval is divided into four pieces, the unit square is divided into $2 \times 2 = 4$ equal squares, and a function analogous to the one in the previous iteration maps each subinterval to a rescaled $\Pi$-shape in each of the subsquares. Some of the shapes are rotated by 90 degrees, and the endpoint of each of them but the last one is connected by a line segment with the initial point of the next S-shape so that they form one piecewise-linear curve. In the next steps, dividing, rescaling, and connecting the endpoints is repeated, multiplying the number of elementary shapes (generators) by 4.

In the limiting case, Peano and Hilbert curves have self-intersections and thus are only surjective, not bijective; however, they are bijective maps of grids at each finite stage of construction, which is sufficient in practical applications. 

An important characteristic of both the Hilbert and Peano curves is Hölder continuity, which establishes a relationship between the distance of the curve $|x-y|$ and the distances of their corresponding curve points $\parallel f(x) -f(y) \parallel$~\cite{bader2012spacefilling}:
\begin{equation}
    \parallel f(x) -f(y) \parallel \leq c_f |x-y|^{1/d}
    \label{eq:Holder_continuity}
\end{equation}
where $\parallel \cdot \parallel$ denotes the Euclidean norm, $d$ is a dimension of the curve, and $c_f$ is a constant. This property makes the SFC effective for quantifying local data properties.

\subsubsection{Data-driven SFC}

The shape of Peano and Hilbert SFCs discussed before is independent of the underlying data. However, if the complexity of the two- or three-dimensional image being analyzed manifestly varies across regions (e.g., local fractal dimension fluctuations), it is reasonable to propose a space-filling curve adaptable to local data characteristics~\cite{Zhou2021SFCDataDriven}. In such a data-driven approach, the length of each subsequent line segment appended to the curve is determined by assessing changes in local pixel values. While this strategy holds promise for potentially enhancing the resulting data series, it has the drawback of being more computationally demanding than applying a curve of a predetermined shape. Nevertheless, it offers the additional possibility of using the curve's length as an estimator for the fractal parameters of the image.

\begin{figure}[H]
    \centering
    \includegraphics[width=0.9\columnwidth]{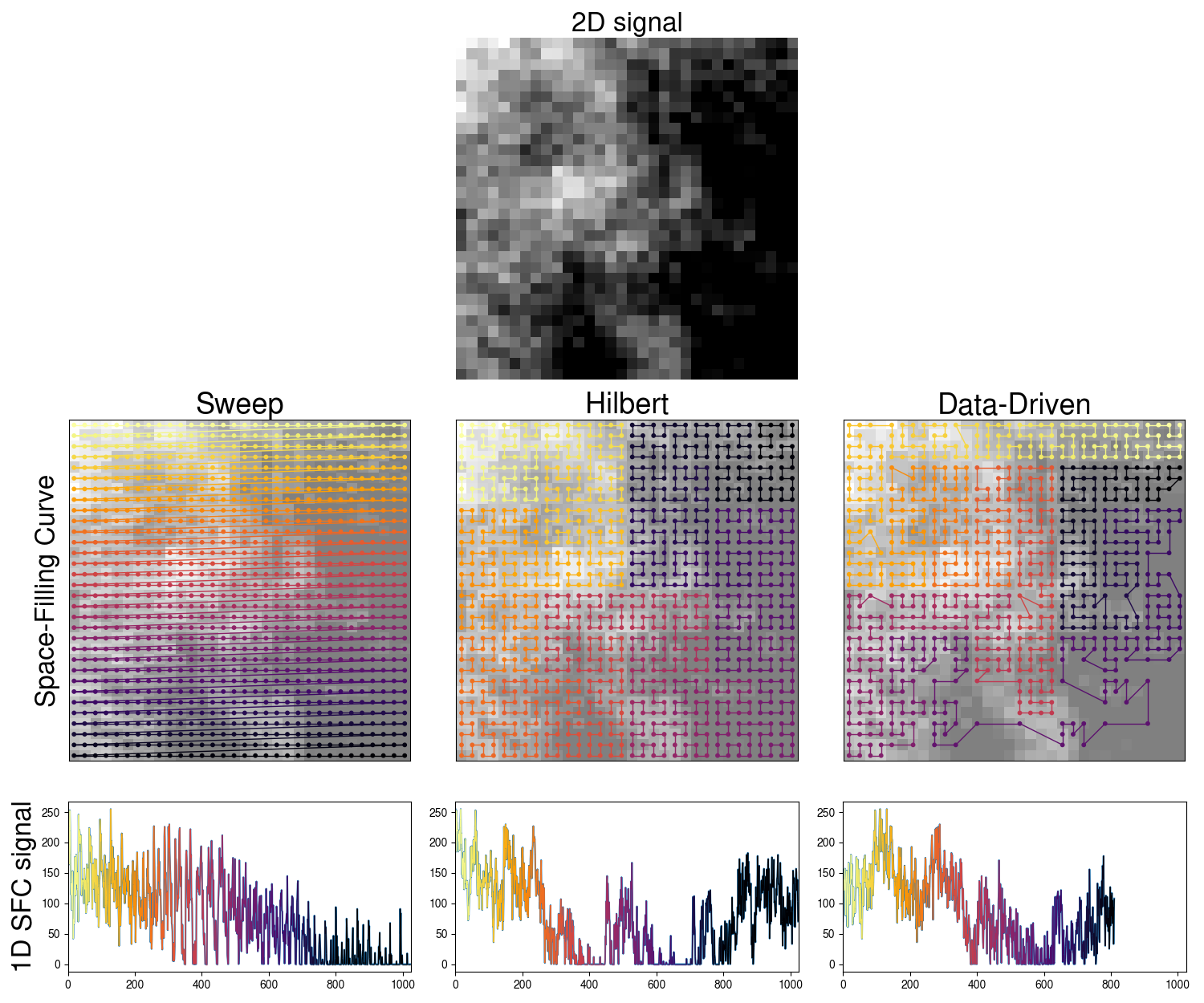}
    \caption{Transformation of two-dimensional data into a one-dimensional series via three types of space-filling curves. Colors code the position of data points in the series (color online).
    }
    \label{fig03:SFCExamples}
\end{figure}

\subsubsection{Non-fractal SFC}

As a benchmark for analyzing the fractal SFC within the FSCA algorithm, we used two simple methods of data linearization: \textit{random}, where all the data entries from the data grid are lined in a random order, and \textit{sweep} (cf. Fig.~\ref{fig03:SFCExamples}), where a two-dimensional data grid is flattened to a one-dimensional list row-by-row.
These simple linearizations do not preserve locality and, in strict mathematical terms, are not well-defined continuous maps from $[0,1] \in \mathbb{R}$ to $ [0,1] \times [0,1] \in \mathbb{R}^2$ in the infinite resolution limit, since two nearby points on the unit interval are not, in general, mapped to two nearby points in the square.
The lack of locality makes these simple procedures less apt than the Hilbert curve to preserve a dataset's correlation properties, as will be demonstrated in later chapters.

\subsubsection{Boundary effects in linearization} 
\label{sec:boundary}

Natural structures and processes occupy geometric spaces that seldom conform perfectly to rectangular data structures. Nonetheless, it has become customary to encapsulate data within tables, and, in fact, the process of dimensional reduction through SFC linearization mandates the utilization of square data grids for which the Hilbert curve is well defined. We propose two methods of addressing this issue, both of which result in boundary effects because of the introduction of artificial data (zeros). Throughout this work, we take care to ensure the robustness of FSCA and test it in the face of such effects.

The input data are embedded within a square by introducing extra space filled with zeros. The length of each side of the square is equal to $2^n$, which is consistent with the length of an $n$-th-iteration Hilbert curve, $n = \log_2 \lceil\max(d_x,d_y,...) \rceil $, where $d_x, d_y,...$ represent the sizes of the input data. We assess the impact of this extra space in both approaches: In the ``padded'' method, the whole square of the embedding is linearized using the Hilbert curve to produce a one-dimensional data sequence. In the ``cropped'' method, a similar procedure is followed, but the additional zeros are marked prior to linearization, and they are eliminated from the resulting time series after linearization.

\subsection{Detrended fluctuation analysis}
\label{DFA}

One of the most dependable methods for detecting and quantifying long-range autocorrelations in a time series is the DFA~\cite{peng1994}. This algorithm 
works even in cases where the time series is non-stationary, rendering it particularly valuable in experimental data analysis and surpassing classical methods. Owing to its effectiveness, DFA finds application across various scientific domains~\cite{KANTELHARDT2001, ochab2022,Grau2001}. The DFA procedure consists of several steps, as described below. 
In the first step, for a time-series $\{x_i\}_{i=1}^{N}$ of length $N$, the profile $Y(j)$ is calculated as
\begin{equation}
Y\left(j\right) =\sum_{i=1}^j\left [x_{i}-\langle x\rangle \right ],
\label{profiles_DFA}
\end{equation}
where $j=1,2...N$, and the angle brackets, $\langle x\rangle$, denote the average over the entire time series.
Next, the profile $Y(j)$ is divided into $N_s = \lfloor N/s \rfloor $ non-overlapping segments of length $s$, where $\lfloor . \rfloor$ denotes the floor function. For both endpoints to be included in the analysis,
the division procedure is performed twice (on the original and the reversed data series)
such that $2N_s$ segments indexed by $\nu$ are formed. To remove possible trends existing in the data, a polynomial $P^{(m)}_{\nu}$ of order $m$ is fitted to each segment and subtracted from the data. In our study, we used $m=2$~\cite{oswiecimka2013}. In the third step, the detrended variance in each segment $\nu$ is calculated as
\begin{equation}
F^{2}(\nu,s)=\frac{1}{s}\sum_{k=1}^{s}\left (Y((\nu-1)s+k)-P^{(m)}_{\nu}(k) \right )^2.
\label{detrended_variance}
\end{equation}
To quantify possible long-range autocorrelations, the average of variances over segments is computed:
\begin{equation}
F_2(s)=\Big\{\frac{1}{2N_s} \sum_{\nu=1}^{2N_s}[F^{2}(\nu,s)]\Big\}^{1/2}. 
\label{Eq_fluctuation_f}
\end{equation}
In the case of long-term dependencies, a power law is observed:
\begin{equation}
F_2(s)\sim s^{H},
\end{equation}
where $H$ denotes the Hurst exponent. For classical Brownian motion, $H = 0.5$. Any deviation from this value indicates an existing correlation in the time series, i.e., for a positively correlated time series (persistent), $H$ assumes values $0.5<H \leq 1$, whereas negative correlation (antipersistence) is quantified by $0 \leq H<0.5$.  
It is worth noting that the correlation length in the time series corresponds to the scaling range of $F_2(s)$, and in the case of data with complicated correlation organization, the exponent in the power law behavior of $F_2(s)$ may depend on the scaling region being considered. Thus, to comprehensively characterize the organization of a time series, one should take care to estimate the exponents in the proper scaling regimes. This aspect requires careful consideration to prove the method's utility to real neuroimaging data. It is also worth noting that estimating the exponent for an integrated time series increases the exponent by one. Thus, for fractional Brownian motion (an integration of colored noise), the exponents estimated by means of DFA will be greater than one.

\section{FSCA analysis of synthetic data}
\label{sec:synthetic}

The first testing ground for the proposed methodology are various computer-generated artificial processes. First, we turn to two-dimensional fractional Brownian motion, a fractal process exhibiting spatial correlation and no temporal dependence. This case corresponds to a two-dimensional slice of an MRI scan. Next, we move to a generalized random Cantor set—a straightforward representation of a binary two-dimensional random process characterized by spatial correlation. This example corresponds to the cortical boundary in two-dimensional slices, as used, e.g., in the computation of gyrification index~\cite{schaer2008ITMISurfacebased}.  Lastly, we investigate a two-plus-one dimensional Gaussian process characterized by correlation in both spatial and temporal dimensions as an example of a process akin to fMRI data. Below, we outline the primary objectives of the analysis.

\begin{itemize}
    \item \textit{Selection of an optimal space-filling curve} to map the data, which is pivotal to our methodology. Consequently, we assessed the performance of commonly used fractal SFCs, confirming their 
    reproducible outcomes. We investigated Hilbert and data-driven SFC, as well as simpler methods such as sweep and random data ordering (cf. Fig.~\ref{fig03:SFCExamples}). We found the Hilbert SFC optimal in terms of computational cost and accuracy.
    \item \textit{Exploring dependence of FSCA on the resolution of the fractal SFC.} 
    Each resolution level represents a different sampling density of the analyzed data. We found that the choice of resolution levels has little impact on the outcomes of the proposed method.
    \item \textit{Testing robustness of FSCA to boundary effects.} 
    When analyzing experimental data, certain unintended effects—such as boundary effects—might impact the results. We address this issue and elaborate on how our methodology remains robust to such effects.
\end{itemize}

The conclusions drawn from this section will serve as a foundation for the analysis of the neuroimaging data presented in the subsequent sections.

\subsection{Spatially correlated process: fractional Brownian motion}
\label{sec:fbm}

One of the fundamental models employed in generating fractal data is the fractional Brownian motion (hereafter referred to as fBm), originally introduced by Mandelbrot and Van Ness~\cite{mandelbrot1968}. This model enables the simulation of trajectories with predefined long-range dependencies and self-similar properties, all governed by a single parameter known as the Hurst exponent. This exponent not only dictates the scaling but also defines the fractal dimension of the process. Due to its straightforwardness, fBm finds widespread application in simulating real stochastic processes characterized by power-law correlations. Formally, fBm serves as a generalization of the classical Brownian motion to encompass scenarios with long-range correlations and with its covariance defined on an $d$-dimensional domain ($W(\vec{x})$ for $\vec{x} \in \mathbb{R}^d$) by:

\begin{equation}
    \left < W(\vec{x}) W(\vec{y})\right > = \frac{1}{2} \left( |\vec{x}|^{2H} + |\vec{y}|^{2H} - |\vec{x}-\vec{y}|^{2H} \right),
\end{equation}
where $H$ signifies the Hurst exponent. In a one-dimensional scenario, the variables $W(\vec{x})$ and $W(\vec{y})$ represent the process values at distinct time points (cf. Section~\ref{DFA})~\cite{barnsley1988fractal}. At $H=1/2$, the classical Brownian motion is recreated. The Hurst exponent is related to the fractal dimension $D$ by the following formula~\cite{gneiting2004hurst,nualart2005brownian,volker2014stochastic}:
\begin{equation}
D = d + 1 - H.
\end{equation}
Thus, the larger the fractal dimension, the smaller the Hurst exponent, and vice versa.
Below, we refer to this underlying Hurst exponent of the images by $H_{\text{image}}$, while we reserve the symbol $H$ for the estimate obtained from DFA of the linearized data.

\begin{figure}[H]
    \centering
    \includegraphics[width=0.32\textwidth]{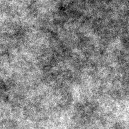}
    \includegraphics[width=0.32\textwidth]{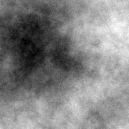}
    \includegraphics[width=0.32\textwidth]{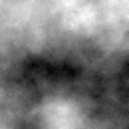}
    \caption{Samples of two-dimensional fractional Brownian motion with different degrees of persistence. \textbf{Left:} antipersistence ($H_{\text{image}}=0.1$, $D=1.9$). \textbf{Middle:} classical Brownian motion ($H_{\text{image}}=0.5$, $D=1.5$). \textbf{Right:} persistence ($H_{\text{image}}=0.9$, $D=1.1$).}
    \label{fig04:Brownian_paths}
\end{figure}

We generate sample paths of the multidimensional fBm with the popular midpoint displacement algorithm (or diamond-square algorithm)~\cite{fournier1982}. Figure~\ref{fig04:Brownian_paths} shows instances of a two-dimensional fBm process generated for three different values of the Hurst parameter. It is easy to notice that for $H_{\text{image}}=0.1$ ($D=1.9$), the image is very rough, with frequent alternations of peaks and valleys, whereas for $H_{\text{image}}=0.9$ ($D=1.1$), it is much smoother. The image with an intermediate value $H_{\text{image}}=0.5$ ($D=1.5$) is an example of a two-dimensional Brownian motion.

\subsubsection{Fractal analysis: Hilbert and data-driven SFCs}

\begin{figure}[H]
    \centering
    \includegraphics[width=\columnwidth]{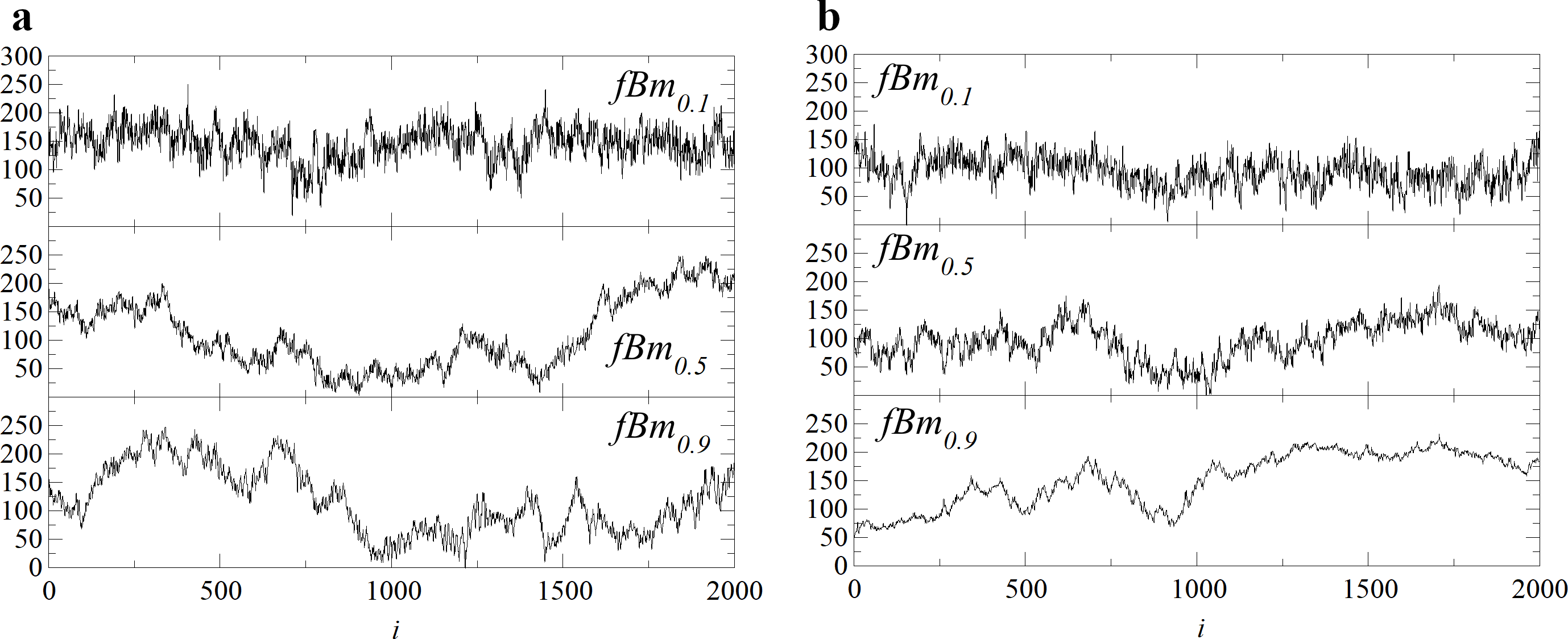}
    \caption{Sample paths of fractional Brownian motion with three different degrees of persistence linearized by two SFCs: a) data-driven, and b) Hilbert.}
    \label{fig05:Brownian2}
\end{figure}

In our investigation, we generate a series of fractional Brownian motion images of size $256 \times 256$ with Hurst exponents $H_{\text{image}}$ ranging from $0.1$ to $0.9$ in increments of $0.1$. For each image, we apply both the Hilbert SFC (at its maximal granularity, conf. \ref{sec:fBMgranularity}) and a data-driven SFC. Subsequently, the resulting data series were subject to analysis using the DFA method. Figure~\ref{fig05:Brownian2} illustrates sample paths of the images corresponding to three distinct regimes of Hurst exponents: antipersistence ($H_{\text{image}}=0.1$), randomness ($H_{\text{image}}=0.5$), and persistence ($H_{\text{image}}=0.9$). Notably, a discernible correspondence exists between the volatility of the data series and the image's level of correlation. 

\begin{figure}[H]
    \centering
    \includegraphics[width=\columnwidth]{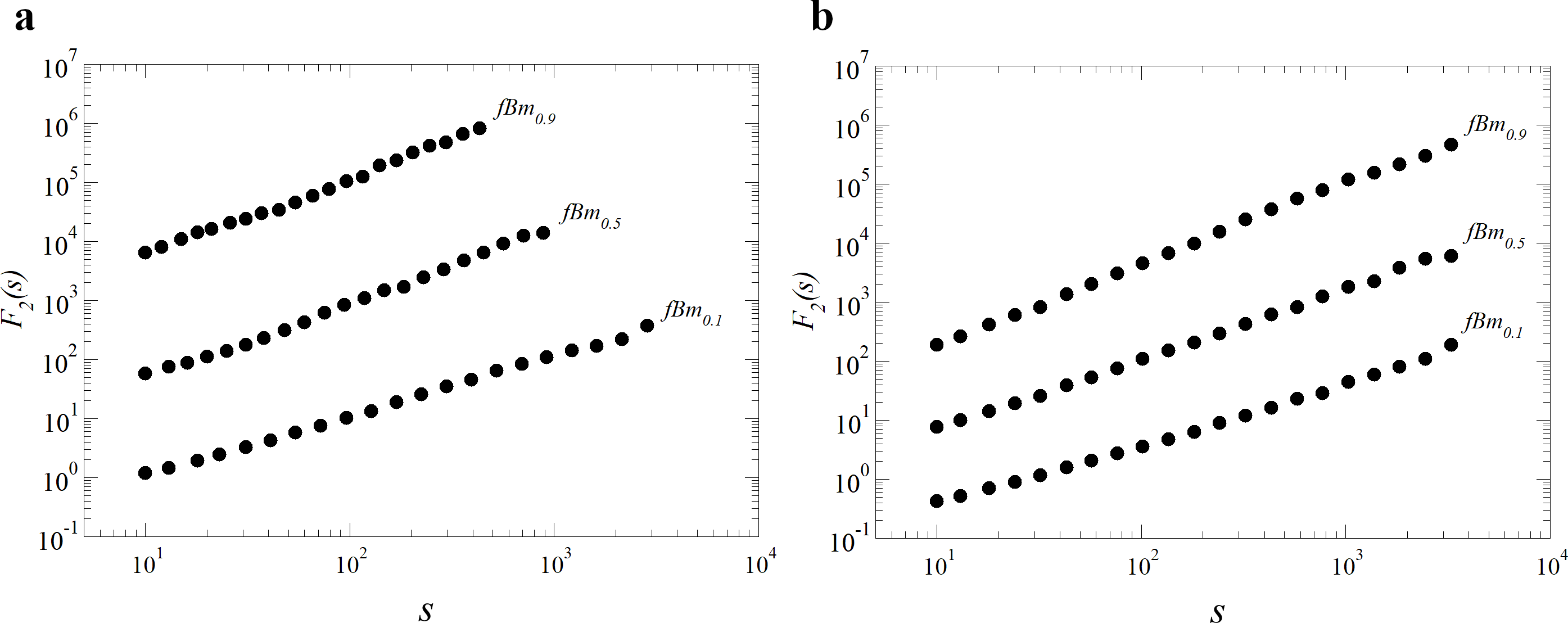}
    \caption{Fluctuation functions $F_2(s)$ calculated for a time series resulting from a fractional Brownian motion linearized using a) data-driven, and b) Hilbert SFCs. The series obtained using data-driven SFC can be shorter than the number of pixels in the images and its length decreases for more persistent images. The Hurst exponents are estimated at two scaling ranges separated at around the image size $s = 250$.}
    \label{fig06:Brownian3}
\end{figure}

For larger Hurst exponents (indicating increased persistence and smaller fractal dimension $D$), the series tend to be smoother, reflecting the spatial organization of the images. It is important to highlight that when employing the Hilbert SFC the resulting series may include periods of low or even zero volatility, coming from highly persistent images. Conversely, the data-driven SFC might skip such segments. Consequently, the series obtained using data-driven SFCs can be shorter than the number of pixels in the images, whereas the series obtained with the application of the Hilbert SFC maintains the original data size.
The effect is visible in Fig.~\ref{fig03:SFCExamples} but not in Fig.~\ref{fig05:Brownian2}, which contains only a fraction of the whole time series. As a result, the range of scales ($s$) utilized for estimating fluctuation functions $F_2(s)$ can be broader for data generated with the Hilbert SFC, cf. Fig.~\ref{fig06:Brownian3} \textbf{a} and \textbf{b}.

\begin{figure}[H]
    \centering
    \includegraphics[width=0.8 \columnwidth]{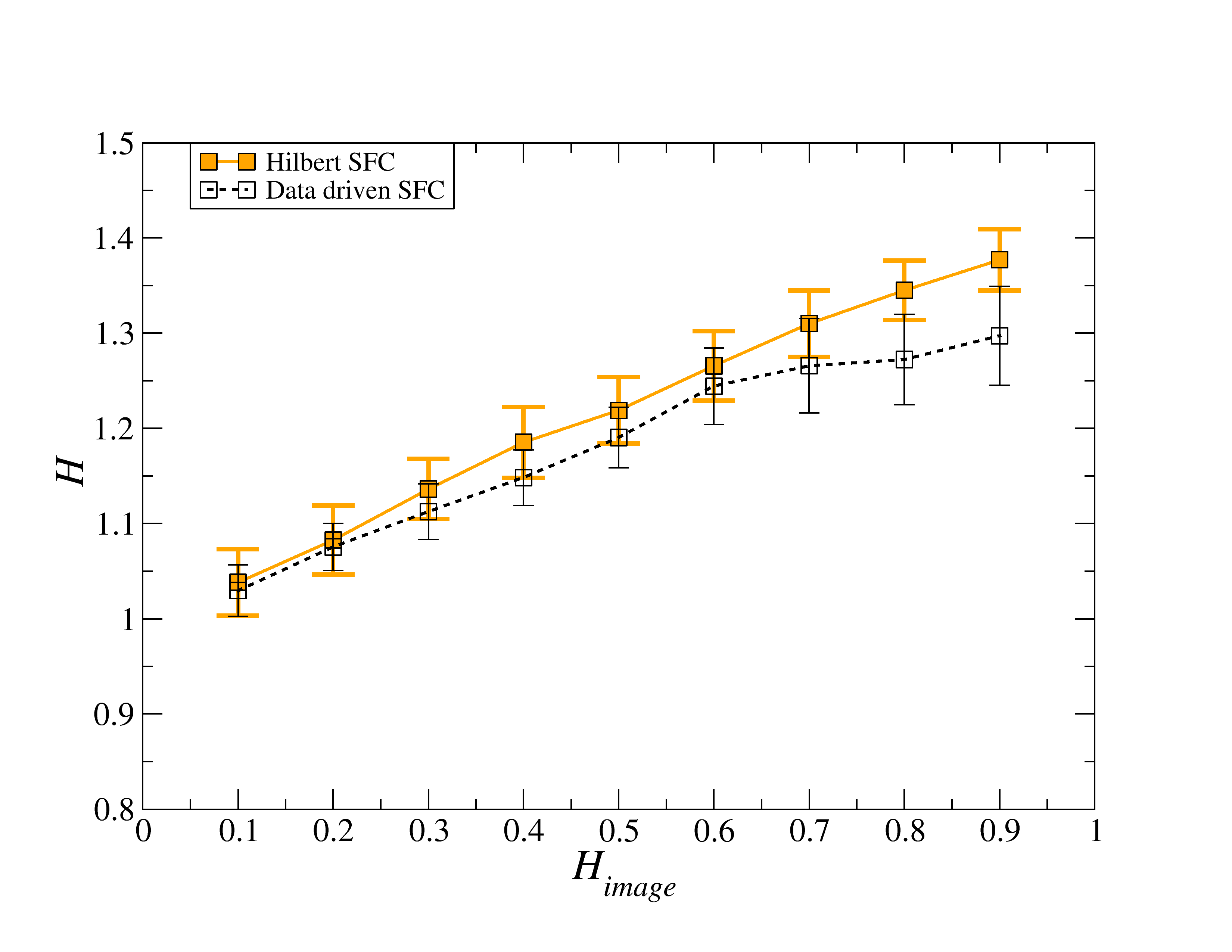}
    \caption{Hurst exponent estimated for data series resulting from a fractional Brownian motion linearized using a Hilbert SFC and a data-driven SFC. Error bars are standard deviations over 100 samples for each value of $H_{\text{image}}$.}
    \label{fig07:Brownian4}
\end{figure}

Figure~\ref{fig06:Brownian3} illustrates the fluctuation functions $F_2(s)$ computed for data series depicted in Fig.~\ref{fig05:Brownian2}. In both cases—Hilbert SFC and data-driven SFC—the function $F_2(s)$ demonstrates nearly perfect power-law scaling for all considered scales, thereby confirming the fractal nature of the analyzed images. Figure~\ref{fig07:Brownian4} summarizes the comparison between two methods of linearization. The clear correlation between the Hurst exponents $H$ obtained through the FSCA with Hilbert SFC and the fBm image exponent $H_{\text{image}}$ is evident, highlighting the high precision of the results due to the broader scale range employed for estimating the fluctuation function. The estimated slope of the linear relationship is approximately 1/2, which may result from the Hölder continuity property of the Hilbert space-filling curve (Eq. \ref{eq:Holder_continuity}). On the contrary, the results obtained through the data-driven SFC show less sensitivity to the spatial organization of the data, leading to a plateau observed in images with high persistence.

\subsubsection{Linearization methods and boundary effects}\label{sec:fBMboundary}

\begin{figure}[H]
    \centering
    \includegraphics[width=\columnwidth]{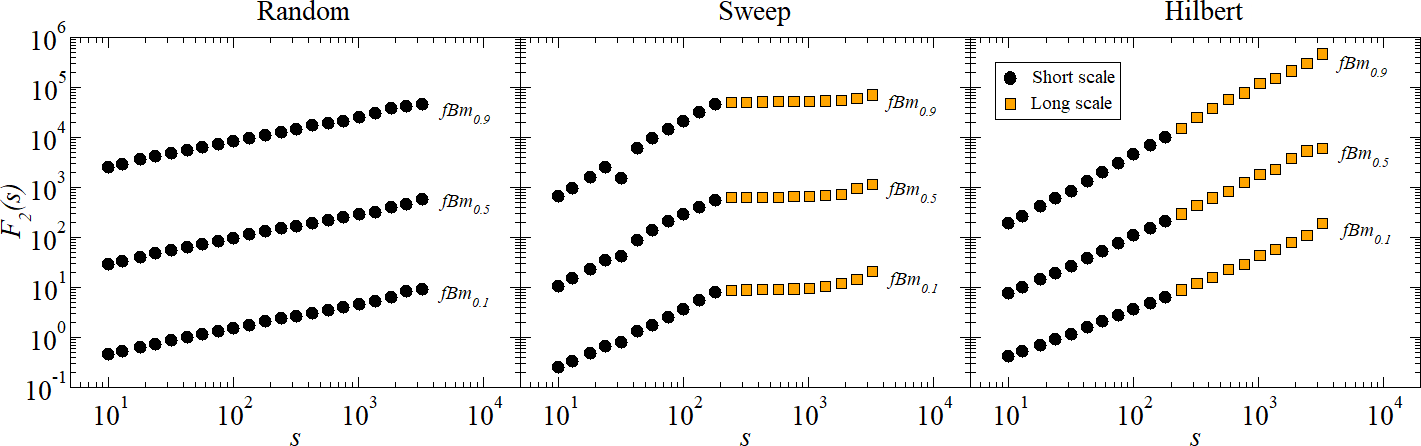}
    \caption{Fluctuation functions $F_2(s)$ calculated for time series resulting from fractional Brownian motion linearized by means of three types of space-filling curves (color online).} 
    \label{fig08:Hilbert2}
\end{figure}

In this section, we test other linearization methods besides Hilbert and data-driven SFCs, as well as take into account the effects resulting from irregularly shaped images.
As in the previous section, we considered the two-dimensional fBM model with different self-similar properties: antipersistent, random, and persistent ($H_{\text{image}}=0.1, 0.5, 0.8$). We mapped the fBM images onto a one-dimensional data series, ordering the pixels according to three linearization algorithms: random ordering, sweep linearization (pixels ordered row by row), and the Hilbert curve. We also considered the impact of the boundary effects – padding and cropped padding (see section \ref{sec:boundary}) – on the obtained final characteristics. The fluctuation functions $F_2(s)$ estimated for the data series resulting from fBm linearized by various pixel orderings are depicted in Fig.~\ref{fig08:Hilbert2}.

\begin{figure}[H]
    \includegraphics[width=1\linewidth]{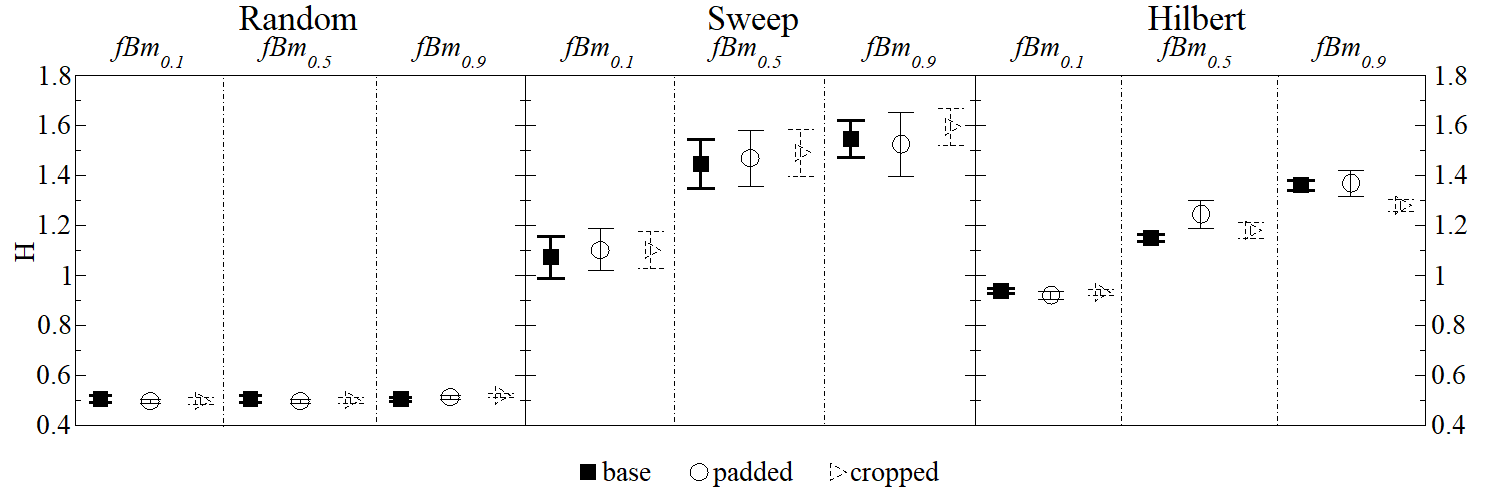}
    \caption{Short-scale Hurst exponent calculated for time series resulting from fractional Brownian motion linearized by means of three types of space-filling curves. Error bars are standard deviations over 10 samples.}
    \label{fig09:brownian5}
\end{figure}

It is evident that the random space-filling curve method produces fluctuation functions that follow a power law; however, the linearization methods determine the differentiation among the processes under consideration. In the case of sweep linearization, the scaling of $F_2(s)$ is observed only at scales below $s=200$, corresponding to the linear size of the image. 
Therefore, with this linearization, the characteristics of the process can only be assessed at these short scales. The Hilbert SFC provides the most reliable insight into the scaling properties of $F_2(s)$ since the power law behavior spans the whole range of available scales.
The estimated Hurst exponents are depicted in Fig.~\ref{fig09:brownian5}. To ensure the stability of our calculations, we present results averaged over ten samples of the process.

From the results of the FSCA method, we can draw the following conclusions: 
\begin{itemize}
    \item When random pixel ordering is applied, the resulting data series is insensitive to the original image pixel intensity correlations. In all cases, the estimated Hurst exponent $H$ equals $0.5$, indicating an uncorrelated process. Also, the boundary effects have no impact on the results. 
    \item The properties of the paths generated by the sweep linearization only exhibit scaling at short scales. The estimated Hurst exponent assumed average values of $H = 1.1, 1.45$, and $1.5$ for the $H_{\text{image}} = 0.1, 0.5$, and $0.8$, respectively. Thus, for the most correlated fBm example, the sweep linearization does not effectively map the properties of the initial process. Moreover, the variance of the results is high (cf. results for the padded data), which decreases precision of this approach.
    \item 
    Hilbert mapping proves to be the most reliable ordering.
    The average exponents of the considered processes assume distinct values $H = 0.9, 1.1$, and $1.4$ for the $H_{\text{image}} = 0.1$, $0.5$, and $0.8$, respectively. The variance is lower than in sweep linearization, and the considered boundary effects have minimal impact on the results. 
\end{itemize}

In summary, our analysis clearly points to the Hilbert SFC as an optimal linearization approach admitting high computational efficiency, good stability of the results, and best preservation of fractal properties of the input data.

\subsubsection{Effects of SFC granularity} 
\label{sec:fBMgranularity}

The non-local character of SFC mappings suggests a granularity study aiming to understand how the ubiquitous real-world issues of data granularity and subsampling will affect the fractal properties found through the proposed method. 
To this end, we apply the usual SFC generation procedure but keep not only the curve obtained in the final iteration step but also several preceding ones, thus obtaining a set of curves whose application to the underlying data results in a set of data series of increasingly finer granularity, interpreted as a higher resolution or better sampling of a real-world data. 

\begin{figure}[H]
    \centering
    \includegraphics[width=\textwidth]{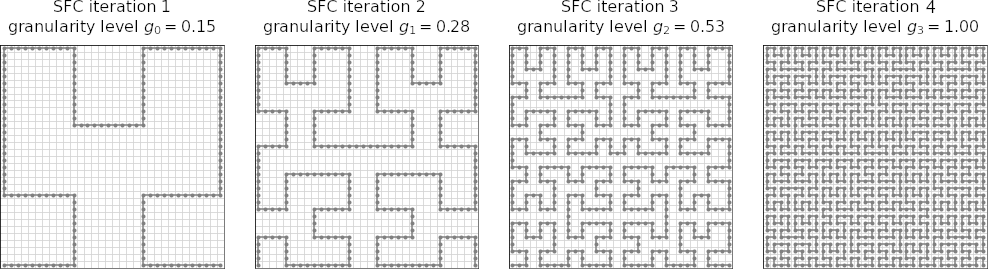}
    \caption{An illustration of the granularity study: Four consecutive iterations of a Hilbert curve with increasingly finer granularity on a square representing a $32 \times 32$ two-dimensional data.}
    \label{fig10:hilbert_granulation}
\end{figure}

In Figure~\ref{fig10:hilbert_granulation}, this methodology is illustrated by four consecutive Hilbert curve iterations. In each iteration, we measure the granularity level \( g_i = \frac{n_i}{n_N} \), which represents the ratio of the \( i \)-th curve length \( n_i \) to the final SFC curve length \( n_N \), where \( i = 0, 1, 2, \ldots, N \) is the iteration number, and \( N \) indicates the final iteration.

\begin{figure}[H]
    \centering
    \includegraphics[width=\textwidth]{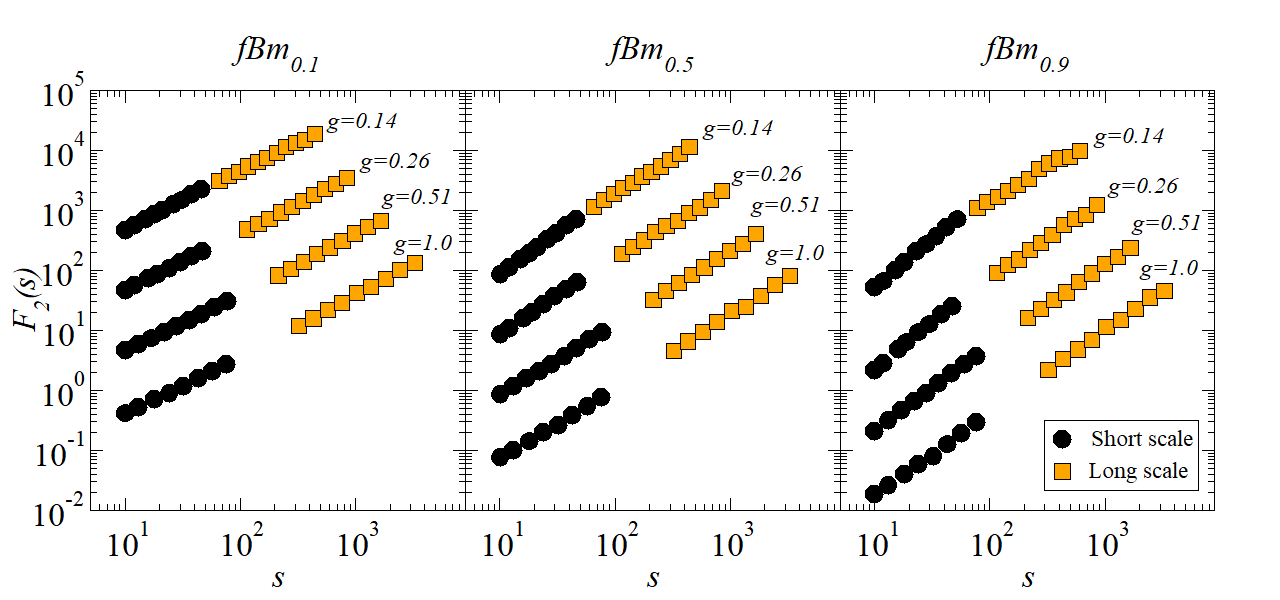}
    \caption{
    Fluctuation functions $F_2(s)$ of a Hilbert SFC curve calculated for the fractional Brownian motion across four granularity levels $g_i$ and three degrees of persistence $H_{\text{image}}$. Only the points in the scaling regimes are shown. (color online)}
    \label{fig11:Hilbert1}
\end{figure}

Despite this significant range of granularity levels, the fractal analysis depicted in Fig.~\ref{fig11:Hilbert1} shows a stable behavior of the Hurst exponents across all four levels of granularity and across all values of $H$ studied.

\subsection{Binary, spatially correlated process: generalized Cantor set}

So far, we argued that the Hilbert SFC mapping is comparable to the data-driven SFC in terms of identifying fractal properties while the former is much faster in terms of computation. However, in the investigation of large data structures, some information compression may be desirable, and it can be achieved by data-driven SFC, which increases its resolution only in the most essential regions. Thus, the number of the curve's segments (equal to the length of the resulting data series) itself provides information about the complexity or compression of the analyzed objects.

We demonstrate this property on the two-dimensional version of the Cantor set called the generalized random Cantor set, which can be generated by an iterated function system (IFS). The generating procedure is as follows. We start with a square, and in the first step, it is divided into $2^{d}$ squares of equal size ($d$ is the topological dimension of the support; $d=2$ in our case), and to each subsquare we randomly assign a number: $1$ with probability $p$ or $0$ with probability $1-p$. In the following steps, the division and number assignment is repeated only for the subsquares marked with $1$, and so on ad infinitum. In practice, we end the process after a finite number of steps, with the final set comprising the squares marked with 1. The fractal dimension $D$ of the generated set is given by
\begin{equation}
D = d + log(p)/log(2). 
\end{equation}

\begin{figure}[H]
    \centering
    \includegraphics[width=\columnwidth]{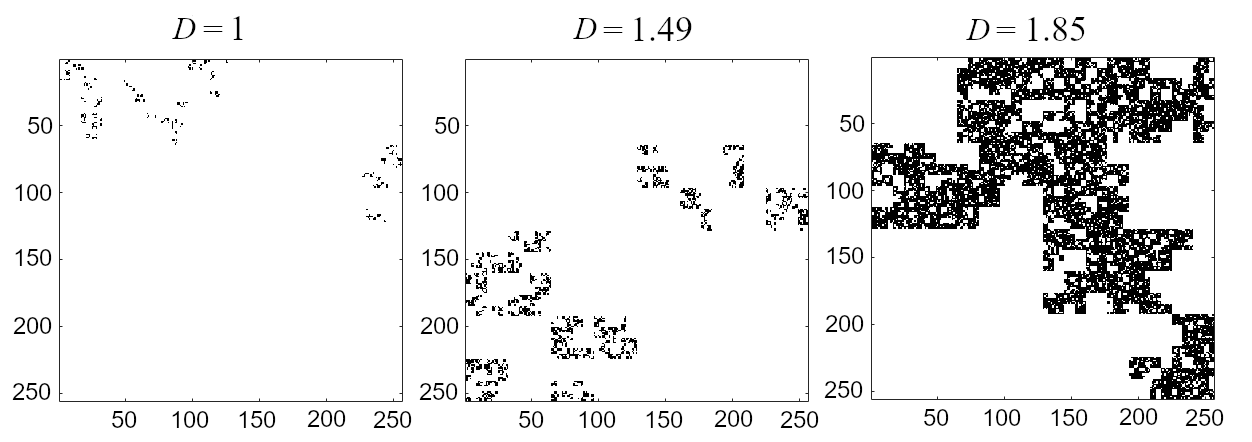}
    \caption{Sample generalized random Cantor sets. Three cases with increasing fractal dimensions $D$ are presented, generated by $p= 0.5$ $(D = 1)$, $0.7$ $(D = 1.49)$, and $0.9$ $(D = 1.85)$.}
    \label{fig12:Cantor_paths}
\end{figure}

\begin{figure}[H]
    \centering
    \includegraphics[width=\columnwidth]{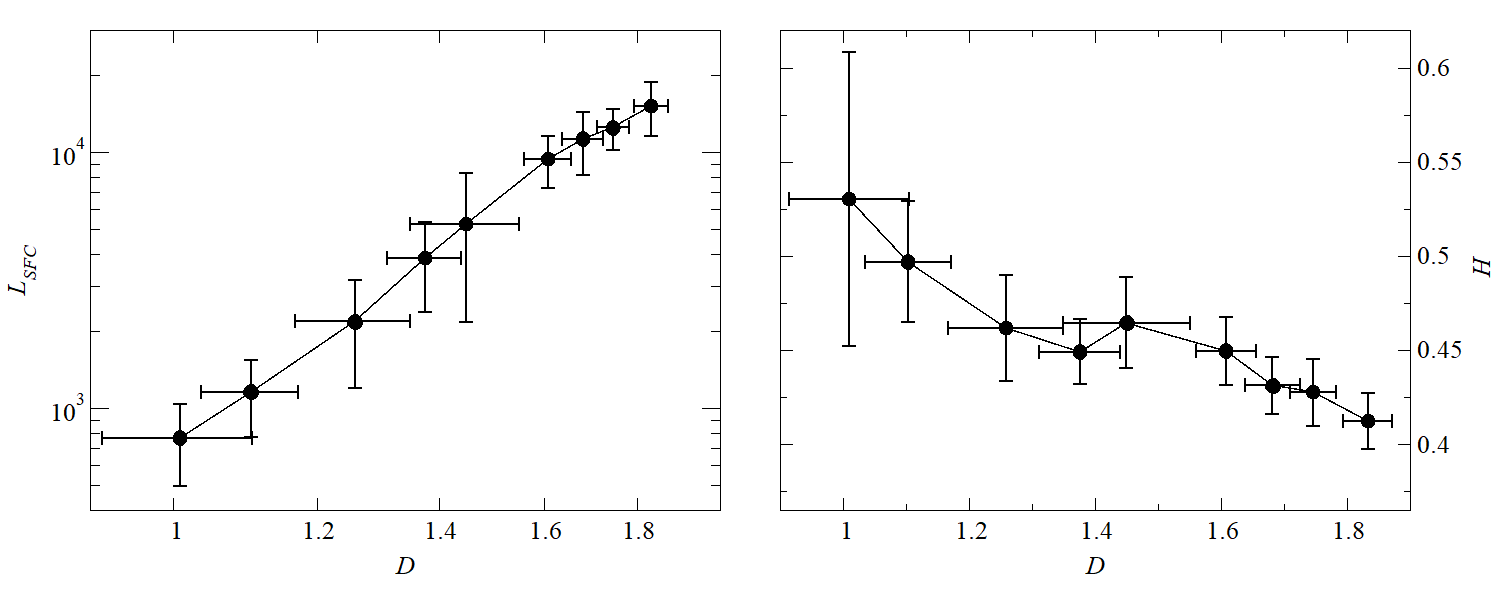}
    \caption{Fractal properties of the generalized Cantor sets measured by FSCA. \textbf{Left:} Relation between the length of the data-driven SFC and fractal dimension $D$ of the two-dimensional Cantor set. \textbf{Right:} Relation between the fractal dimension $D$ of the two-dimensional Cantor set and the Hurst exponent estimated for time series after data-driven SFC linearization.}
    \label{fig13:Cantor_Hurst}
\end{figure}

In our study, we generated random Cantor sets of size $256 \times 256$ and with $p$ in the range $[0.1,0.9]$ with step $0.1$. Sample paths of the processes are shown in Fig.~\ref{fig12:Cantor_paths}. We applied a data-driven SFC procedure for each case and estimated the length of the linearized series. The results are depicted in Fig.~\ref{fig13:Cantor_Hurst}. As expected, the time series length, $L_{SFC}$, increases with the fractal dimension, reflecting the complexity of the data. Moreover, the relation between $D$ and $L_{SFC}$ is exponential, which is easy to notice from the linear relation on the semi-logarithmic plot. The estimated Hurst exponent decreased as the fractal dimension $D$ increased, indicating slight antipersistence for most fractal dimensions. What is worth noticing is that the statistical reliability of the estimated $H$ strongly depends on the fractal dimension, i.e., a smaller fractal dimension denotes a shorter time series and at the same time a lower reliability of the results.  

\begin{figure}[H]
    \centering
    \includegraphics[width=0.8 \columnwidth]{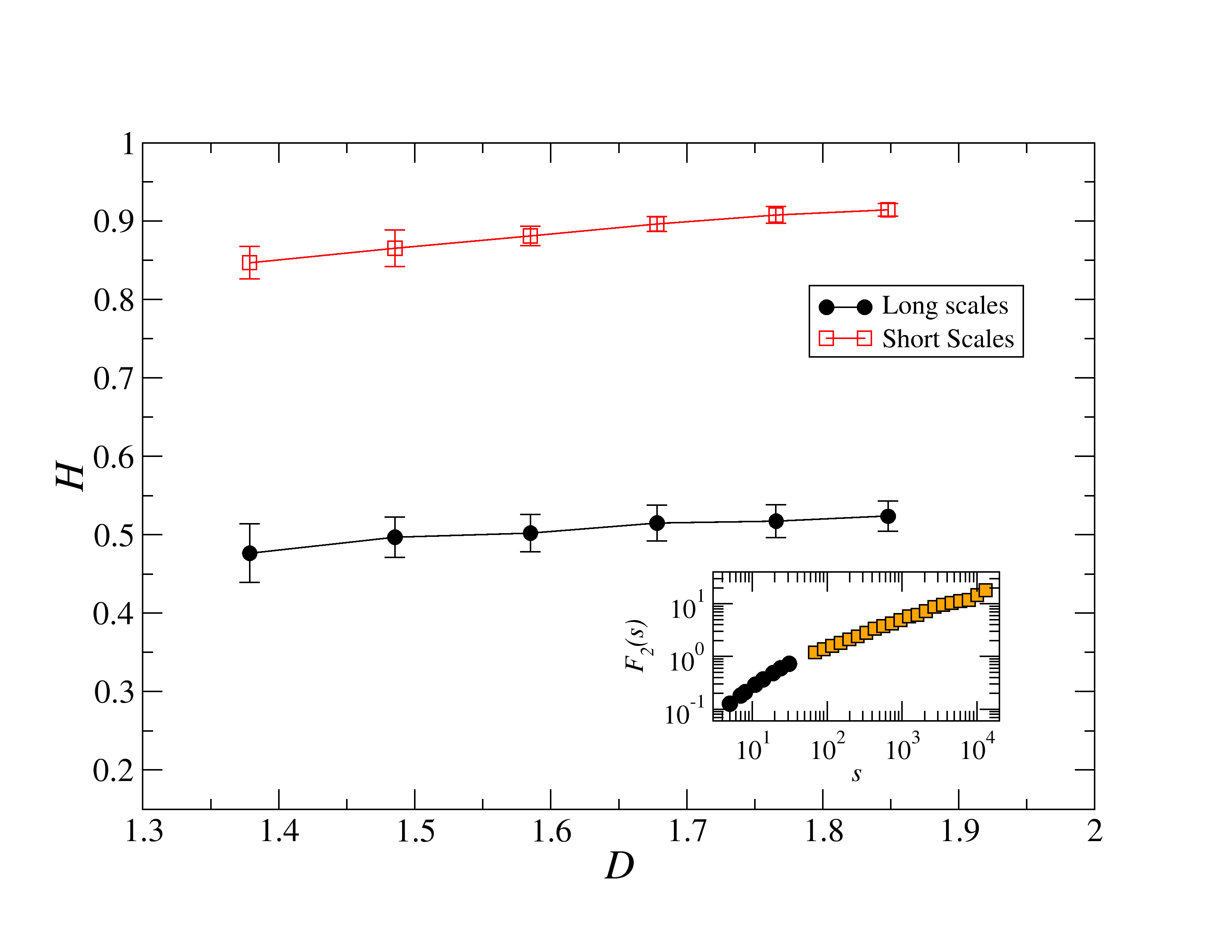}
    \caption{Hurst exponent estimated for data series resulting from generalized random Cantor set.Main: Hurst exponent estimated via FSCA with Hilbert SFC as a function of the fractal dimension $D=$1.85. Inset: Sample fluctuation function for the Cantor set with  Colors indicating different scaling regimes. Error bars are standard deviations over 100 samples.}
    \label{fig14:Cantor_Hurst}
\end{figure}

We also applied the FSCA with Hilbert SFC to analyze the Cantor sets with the fractal dimension $D$ ($D \geq$ 1.38, $p\geq $0.65). The analysis results are depicted in Fig.~\ref{fig14:Cantor_Hurst}. In this case, two scaling regimes were identified; thus, Hurst exponents were estimated independently for each regime. We observe a weak relation between the Hurst and fractal dimension D that is clearer for larger scales than small ones, which stems from the uniform distribution of the measure on the Cantor set and results in the small volatility of the time series.    

\subsection{Spatio-temporal random process: Gaussian process}
\label{sec:gp}

In Section \ref{sec:fbm}, we studied fractional Brownian motion as a paradigmatic example of a correlated random process, here in two dimensions interpreted as space but not in time. Now we focus on processes correlated in $d+1$ dimensions, with one distinct dimension interpreted as time. A broad framework for such phenomena is captured by the Gaussian process~\cite{Rasmussen2005GPBook} $W(\vec{x},t)$, described by a covariance function $K$:
\begin{equation}
    \left < W(\vec{x},t) W(\vec{y},s) \right > = K(\vec{x},t;\vec{y},s)
\end{equation}
and the process mean $\left < W(\vec{x},t) \right > = 0$ which we set to zero.

\begin{figure}[H]
    \centering
    \includegraphics[width=0.24\textwidth]{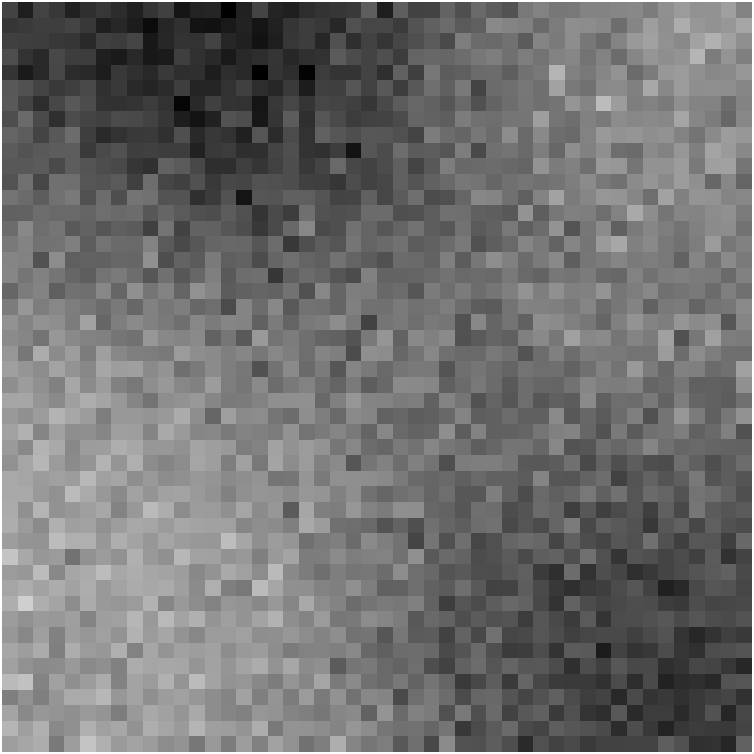}
    \includegraphics[width=0.24\textwidth]{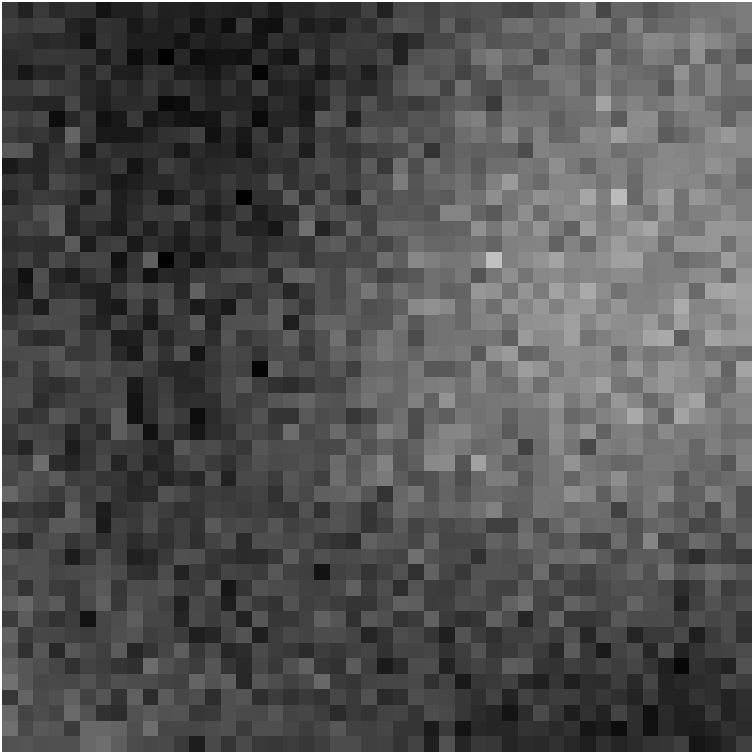}
    \includegraphics[width=0.24\textwidth]{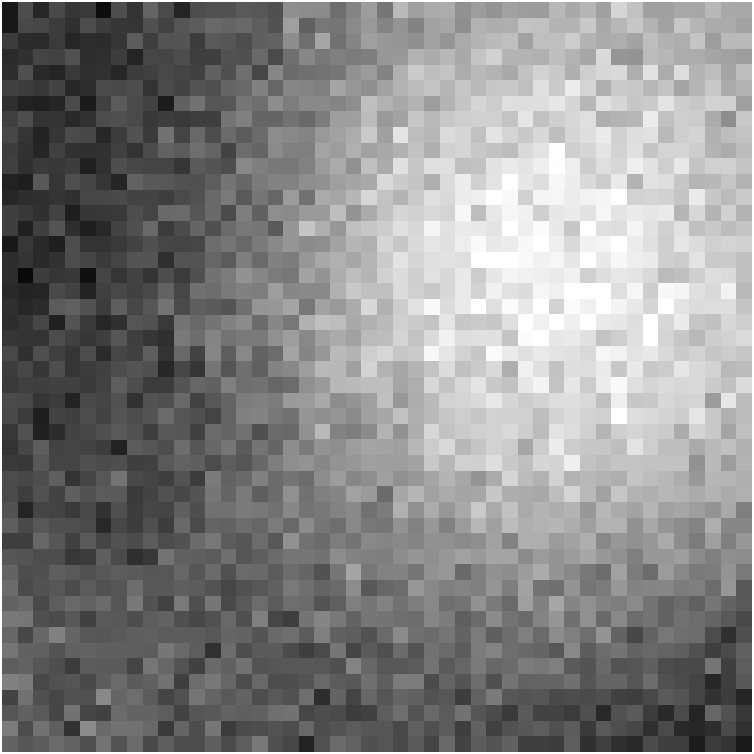}
    \includegraphics[width=0.24\textwidth]{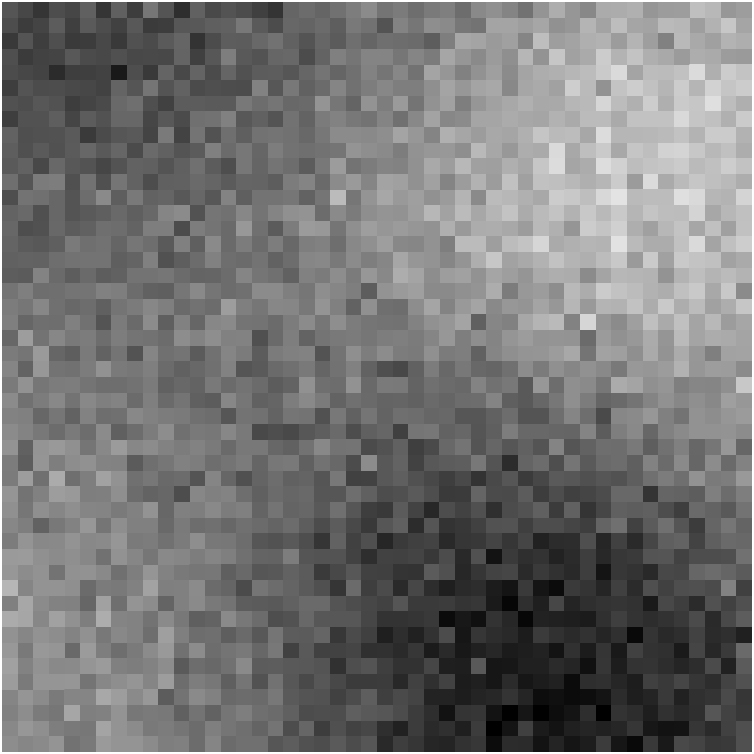}
    \caption{An instance of a Gaussian process with two fixed activity sources located at \( \vec{s}_0 = (0.2,0.4) \) and \( \vec{s}_1 = (0.9,0.8) \), depicted at time steps $1$, $10$, $20$ and $30$. }
    \label{fig14:Hilbert3}
\end{figure}

To maintain continuity with our previous analysis, we focus on a $2+1$-dimensional process. We employ a factorized exponential covariance function, expressed as $K(\vec{x},s; \vec{y},t) = K_{S}(\vec{x},\vec{y}) K_{T}(s,t)$, where the spatial covariance function is defined as:
\begin{align*}
K_{S}(\vec{x},\vec{y}) & = e^{-\frac{|\vec{x}-\vec{s}_0|^2}{2l_s^2}} + e^{-\frac{|\vec{y}-\vec{s}_0|^2}{2l_s^2}} + e^{-\frac{|\vec{x}-\vec{s}_1|^2}{2l_s^2}} + e^{-\frac{|\vec{y}-\vec{s}1|^2}{2l_s^2}},
\end{align*}
while the temporal covariance function is given by $K{T}(s,t) = e^{-\frac{|s-t|}{l_t}}$. The process is characterized by two correlation lengths: \( l_s = 0.4 \) and \( l_t = 0.1 \), along with two fixed activity sources located at \( \vec{s}_0 = (0.2,0.4) \) and \( \vec{s}_1 = (0.9,0.8) \). In Figure~\ref{fig14:Hilbert3}, we present four snapshots illustrating the dynamics of the generated Gaussian process. We clearly identify activity sources at the upper-right and the lower-left corners with sizes proportional to the spatial $l_s$ scale. Simultaneously, the sources also decay in time due to temporal correlations parametrized by $l_t$. The simulation was executed on a \( 48 \times 48 \) grid over \( 100 \) time steps. 

\subsubsection{Fractal analysis: application of SFC}

In this case, we apply the SFC linearization at each time step independently and report the time-averaged fractal properties. We repeat the analysis conducted for the fBm and reapply three linearization maps preserving fractal properties to a varying degree. 


\begin{figure}[H]
    \centering
    \includegraphics[width=\textwidth]{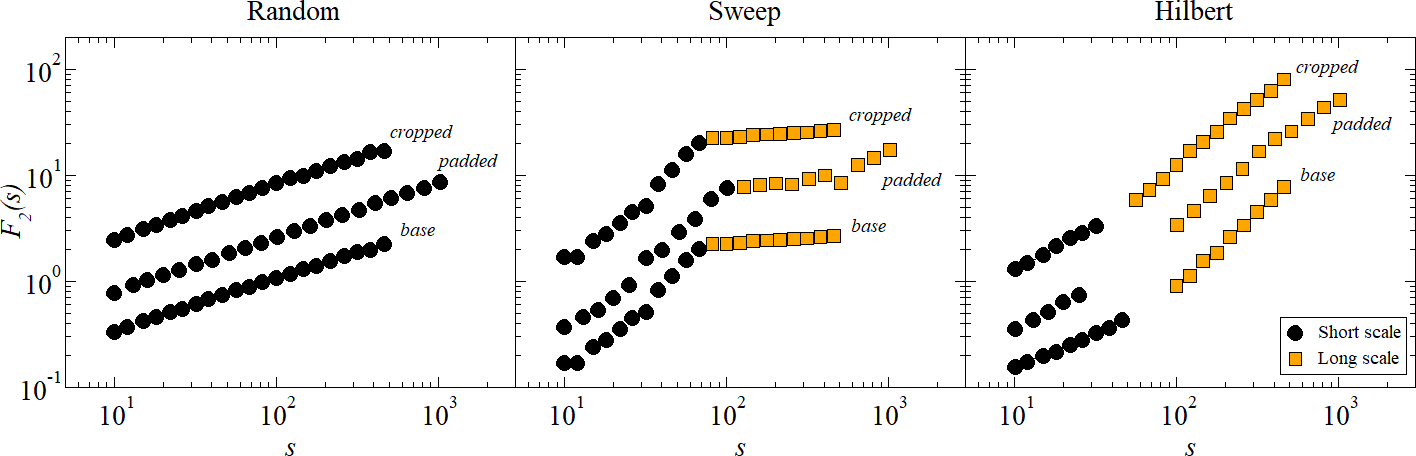}
    \caption{Time-averaged fluctuation functions $F_2(s)$ of the Gaussian process with random, sweep, and Hilbert SFC linearizations, and for three boundary effect scenarios: base, padded, and cropped. For the Hilbert SFC and sweep ordering, short- and long-scale regimes are marked by black circles and yellow squares respectively (color online). The fluctuation functions are averaged over 100 time steps.}
    \label{fig15}
\end{figure}

Figure~\ref{fig15} shows the time-averaged fluctuation function \(F_2\) across different scales \(s\) for the three types of linearizations. The random ordering (left) completely disrupts correlations within the data, with $F_2(s)$ manifesting as a straight line with a constant slope \(H=0.5\), indicative of a white-noise signal. The sweep linearization (center) maintains fractal properties at short scales ($H_{\text{short}}$ = $1.3$, $1.3$, and $1.32$ in cropped, padded, and base scenarios) up to approximately \(s \approx 100\), corresponding to the grid's linear size. Finally, the Hilbert SFC (right) exhibits a scale-dependent behavior: $H_{\text{short}} = 0.82$ (cropped), $0.82$ (padded) and $0.66$ (base) at short scales and $H_{\text{long}} = 1.25$ (cropped), $1.2$ (padded), and $1.42$ (base) at long scales.

Compared with the fBm case illustrated in Fig.~\ref{fig08:Hilbert2}, the fluctuation functions for random and sweep orderings take on similar shapes. In contrast, \(F_2\) for the Hilbert SFC shows a crossover between short- and long-scale Hurst indices, which was absent in the fBm case.

\subsubsection{Boundary effects and SFC granularity}


We revisit the granularity and boundary effects described in Secs.~\ref{sec:fBMboundary}–\ref{sec:fBMgranularity}. For each scenario with three types of linearization, the boundary effects are illustrated in Fig.~\ref{fig15}. We pad the square of size $48\times 48$ with additional zeros to a final square $64 \times 64$ fitting the $n = 6$ generation Hilbert curve.

The findings are consistent with the fBM case, and the resulting Hurst exponents are consistent across all the padding types, confirming the robustness of the method against boundary effects.

\begin{figure}[H]
    \centering
    \includegraphics[width=0.4\textwidth]{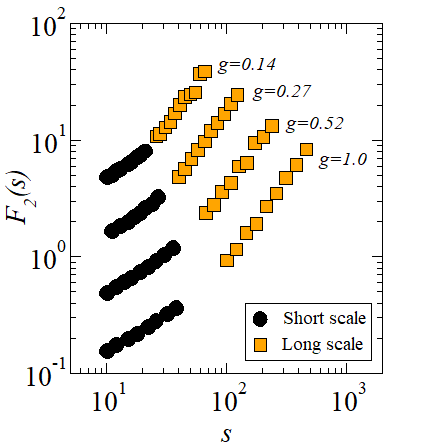}
    \caption{Time-averaged fluctuation function $F_2(s)$ for the Gaussian process processed with Hilbert SFC at four granularity levels. Short- and long-scale regimes are denoted by black circles and orange squares respectively (color online). The fluctuation functions are averaged over 100 time steps.}
    \label{fig16}
\end{figure}

The effects of SFC granularity on FSCA results for the Gaussian process are presented in Fig.~\ref{fig16}. We apply four Hilbert SFCs of increasing granularity levels. Both short- and long-scale Hurst exponents are consistent across all four granularity levels, providing stronger evidence for FSCA robustness.



\subsubsection{Temporal changes of the Hurst exponent}

\begin{figure}[H]
    \centering
    \includegraphics[width=\textwidth]{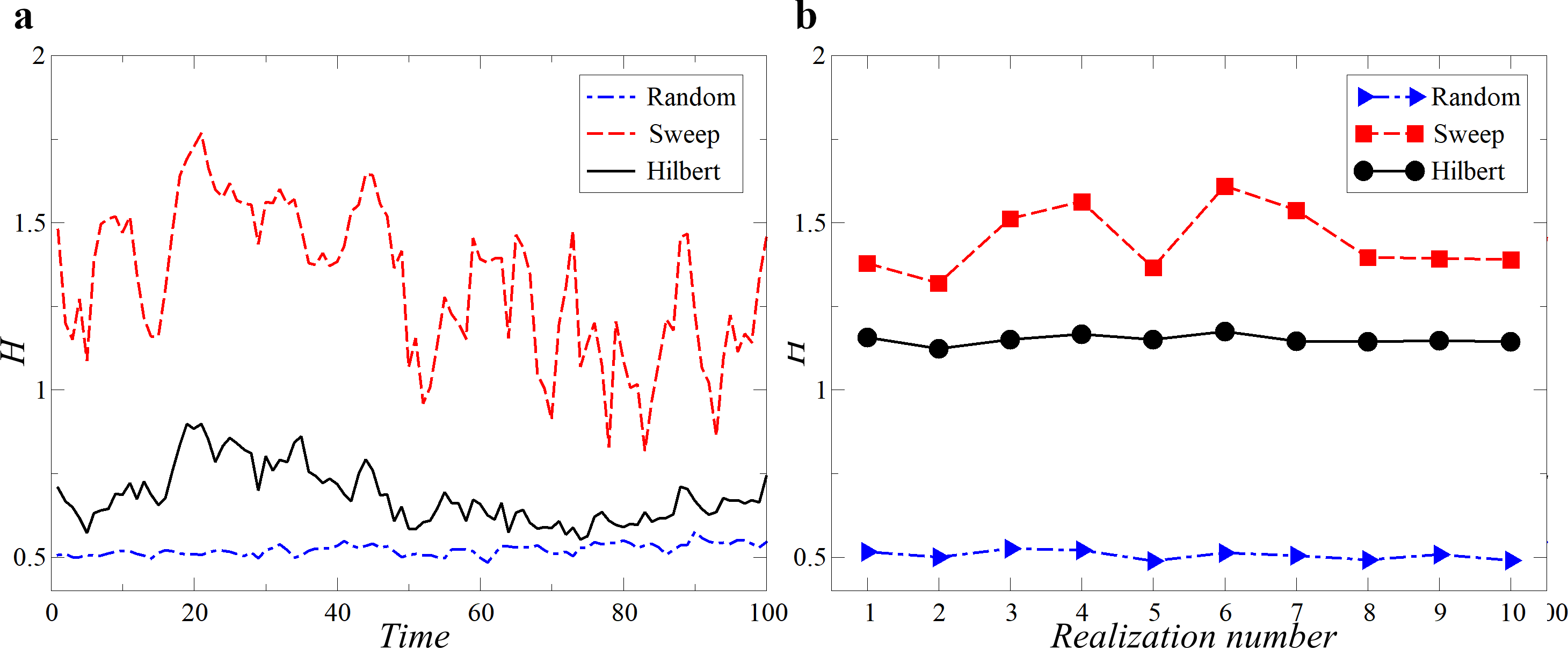}
    \caption{Comparison of time dynamics of the Hurst exponent between a) fractional Brownian motion and b) Gaussian process for random, sweep, and Hilbert SFC linearizations.}
    \label{fig17}
\end{figure}

Thus far, we have focused on time-averaged quantities; in Fig.~\ref{fig17}, we explore the temporal dynamics of a Gaussian process and make a comparison with the non-temporal fractional Brownian motion.  We plot time-resolved Hurst exponents for the three types of SFCs for the fBM process (left plot) and for the Gaussian process (right plot). Across both processes, corresponding Hurst exponents have similar general characteristics. In both cases, we observe a general ordering $H_{\text{random}} < H_{\text{Hilbert}} < H_{\text{sweep}}$. Unsurprisingly, the Hurst exponent processed with the random SFC is centered around \(H \approx 0.5\) for both processes. Clear differences are observed in the sweep-SFC-processed and Hilbert-SFC-processed Hurst exponents which admit stationary behavior (fBM process) or show temporal dynamics (Gaussian process). Although both sweep and Hilbert SFCs follow the time dynamics of the Gaussian process, the latter has a decidedly lower variability across time. This observation is corroborated by the variability in the fBm case reported in Fig.~\ref{fig09:brownian5} across different persistence values. 

\section{Application of FSCA to neuroimaging data}
\label{sec:neuroimaging_data}

We utilize the FSCA method to examine neuroimaging data and validate the effectiveness of the methodology in practical scenarios. All analyses in the following chapter are performed with the Hilbert SFC at maximal granularity level. Our analysis encompasses both structural and functional magnetic resonance imaging datasets. Specifically, we employ the FSCA to scrutinize datasets obtained from individuals diagnosed with Alzheimer's disease. To this end, we draw upon datasets from the Alzheimer's Disease Neuroimaging Initiative (ADNI)~\cite{ADNI} and the Open Access Series of Imaging Studies (OASIS)~\cite{OASIS1}. To study dynamic alterations in the brain's fractal properties, we examine the Poldrack dataset of single-subject, task-oriented fMRI scans~\cite{Poldrack2015}.

\subsection{Structural MRI data analysis: Alzheimer's disease}

Alzheimer's disease (hereafter AD) is a neurodegenerative condition impacting over 50 million individuals globally~\cite{breijyeh2020}, recognized as the predominant cause of dementia. While diagnosing severe stages of AD is straightforward, identifying its early stages poses a significant challenge~\cite{porsteinsson2021}. Among the conventional diagnostic tools employed, magnetic resonance imaging (MRI) stands out, as it provides detailed anatomical images of the brain. In this paper, we use FSCA to analyze MRI datasets obtained from patients in the advanced and early stages of AD. We implement standard preprocessing procedures for every MRI scan, resulting in a three-dimensional data volume of $182 \times 218 \times 182$ voxels. Each voxel encapsulates $1~\text{mm}^3$ of brain matter volume.
We apply FSCA to each slice independently along the $x$, $y$, and $z$ axes (e.g. see Supplementary Fig. 1) 
and conduct the final fractal analysis.
This procedure generates a list of Hurst exponents that vary along the sliced axis (hereafter called the Hurst profile), offering valuable and interpretable insights into the local fractal properties of the brain.


\subsubsection{Description of the analyzed datasets: ADNI and OASIS}
The Alzheimer's Disease Neuroimaging Initiative (ADNI) database (\url{adni.loni.usc.edu}) is a repository, 
whose primary goal has been to test whether serial magnetic resonance imaging, positron emission tomography, other biomarkers, and clinical and neuropsychological assessment can be combined to measure the progression of mild cognitive impairment (MCI) and early Alzheimer’s disease (AD).
It encompasses data from both healthy individuals and those diagnosed with MCI or AD, thus serving as an invaluable resource in the advancement of early detection methods and the development of therapeutic strategies
For up-to-date information, see \url{www.adni-info.org}.

\begin{table}[h!]
    \centering

\begin{tabular}{c|c|c|c| c|c|c |c |c}
    dataset & property & yNC & eNC & EMCI & MCI & LMCI & AD & total \\
    \hline
    \multirow{2}{*}{ADNI} & \# patients & 0 & 36 & 29 & 6 & 24 & 34 & 129\\
     & mean age & - & 73.4 & 71.4 & 68.7 & 71.1 & 72.3 &  \\
     \hline
    \multirow{2}{*}{OASIS-1} &\# patients & 201 & 135 & 70 & 28  & 0 & 2 & 436 \\
     & mean age & 26.8 & 69.1 & 76.2 & 77.8 & - & 82.0 & \\

     \hline
     \hline
    dataset & property & \multicolumn{3}{|c|}{MCI\_NOT\_AD}  & \multicolumn{3}{|c|}{MCI\_AD} & total \\
    \hline
    \multirow{3}{*}{OASIS-3} & time-span [yr] & \multicolumn{3}{|c|}{1/2/3/4/5}  & \multicolumn{3}{|c|}{1/2/3/4/5} & \\
      &\# scans & \multicolumn{3}{|c|}{48/33/21/13/11} & \multicolumn{3}{|c|}{19/22/31/16/14} & 228 \\
     & mean age & \multicolumn{3}{|c|}{76.3} & \multicolumn{3}{|c|}{75.5} & 
\end{tabular}
    \caption{Mean age and cohort size for ADNI and OASIS datasets used in the study. Patient group abbreviations: yNC -- young, normal cognition, eNC -- elderly, normal cognition, EMCI -- early-stage mild cognitive impairment, MCI -- mild cognitive impairment, LMCI -- late-stage mild cognitive impairment, AD -- Alzheimer's disease, MCI\_NOT\_AD -- MCI not progressed to AD within a fixed time-span MCI\_AD -- MCI progressed to AD within a fixed time-span. The OASIS-3 dataset is grouped into time-spans of 1-5 years.}
    \label{tab:cohort_sizes}
\end{table}

Our study considers a subset of 129 MRI scans extracted from ADNI-1 and ADNI-2 studies. The dataset consists of individuals grouped into four cohorts reflecting disease progression, including a healthy control group (NC), early- and late-stage Mild Cognitive Impairment groups (EMCI and LMCI), and an Alzheimer's Disease group (AD). Basic information on group sizes is given in Tab. \ref{tab:cohort_sizes}.

The Open Access Series of Imaging Studies (OASIS) is another large-scale longitudinal neuroimaging investigation focused on understanding and characterizing the aging process in the human brain. It involves the acquisition of structural and functional brain imaging data from a large cohort of individuals over time to explore changes in brain structure and function associated with aging. It was made freely accessible to the scientific community. It encompasses magnetic resonance data covering aging and AD across a diverse demographic, cognitive, and genetic spectrum.

In our study, we concentrate on the OASIS-1 database, which comprises a collection of 416 MRI sessions organized into four cohorts based on the Clinical Dementia Rating (CDR) score (Young Control without CDR score, Elderly Control with $\text{CDR} = 0$, EMCI with $\text{CDR} = 0.5$, and MCI with $\text{CDR} = 1$)~\cite{CDR}. Basic information on group sizes is given in Tab. \ref{tab:cohort_sizes}. Given the well-defined stages of Alzheimer's within this database, we conducted a study on disease progression with the proposed FSCA method.

Lastly, we utilize a subset of the longitudinal OASIS-3 database comprising of 228 MRI sessions of MCI-diagnosed patients with $\text{CDR} = 0.5$ or $1.0$. Patients are grouped based on whether or not they develop dementia within a fixed time span (1-5 years), designated as MCI\_AD and MCI\_NOT\_AD, respectively. Basic information on group sizes, categorized by dementia progression and time span is provided in Tab. \ref{tab:cohort_sizes}. This subset of data serves as a clinically relevant evaluation of the proposed FSCA method for detecting early-stage MCI patients likely to progress to dementia.

\subsubsection{ADNI dataset: Hurst profile of Late-Stage Alzheimer's Disease}
\label{sec:ADNIhurst}

At the outset, we apply the FSCA method to MRI scans from both the control group NC (36 subjects) and the AD group (34 subjects).

As previously stated, we linearized each MRI slice along the $x$, $y$, and $z$ axes, subsequently computing the Hurst exponent for the resulting time series. Additionally, our preliminary calculations had suggested that the data under scrutiny exhibit two distinct scaling regimes. 

\begin{figure}[H]
    \centering
    \includegraphics[width=0.51\textwidth]{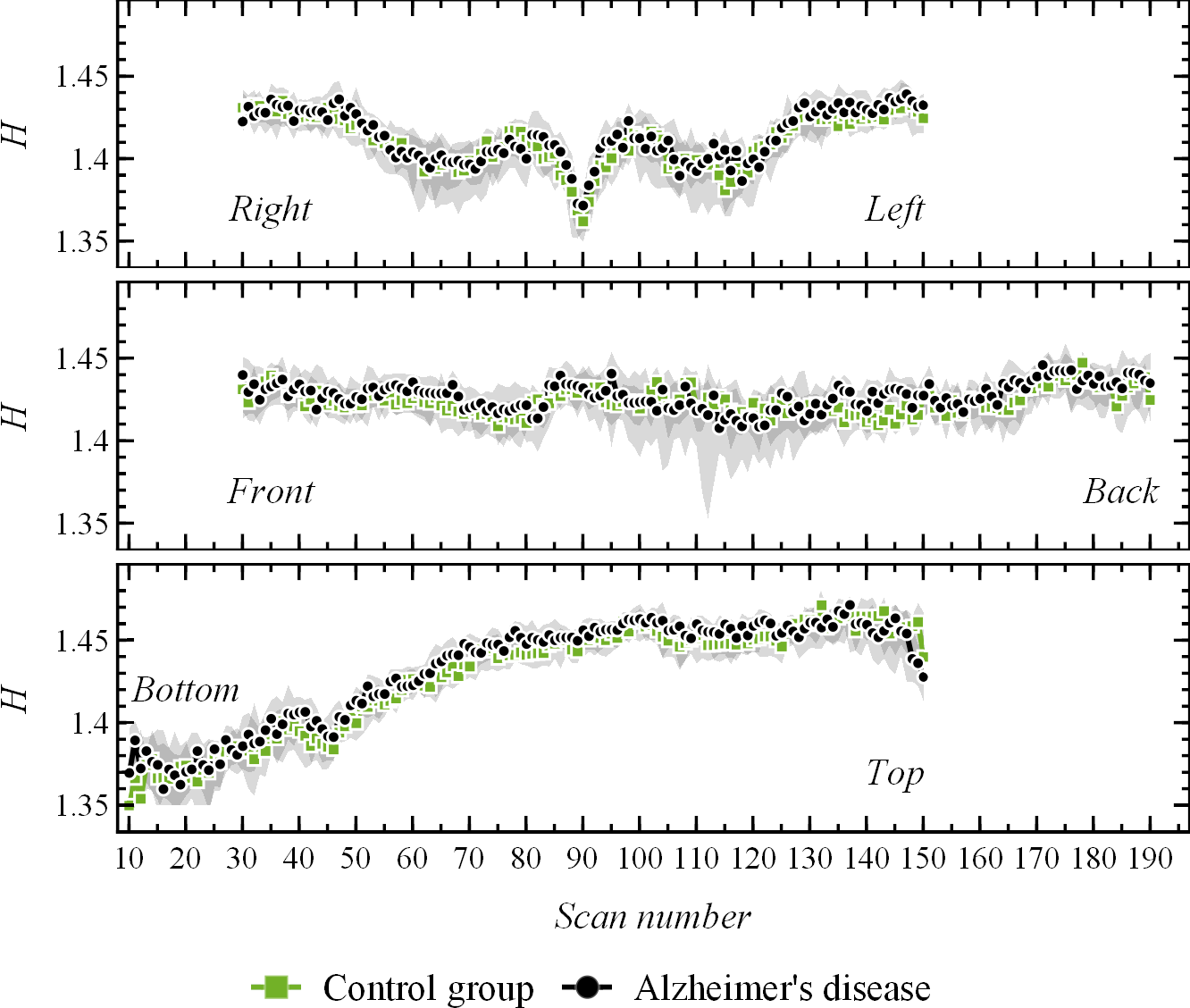}
    \includegraphics[width=0.48\textwidth]{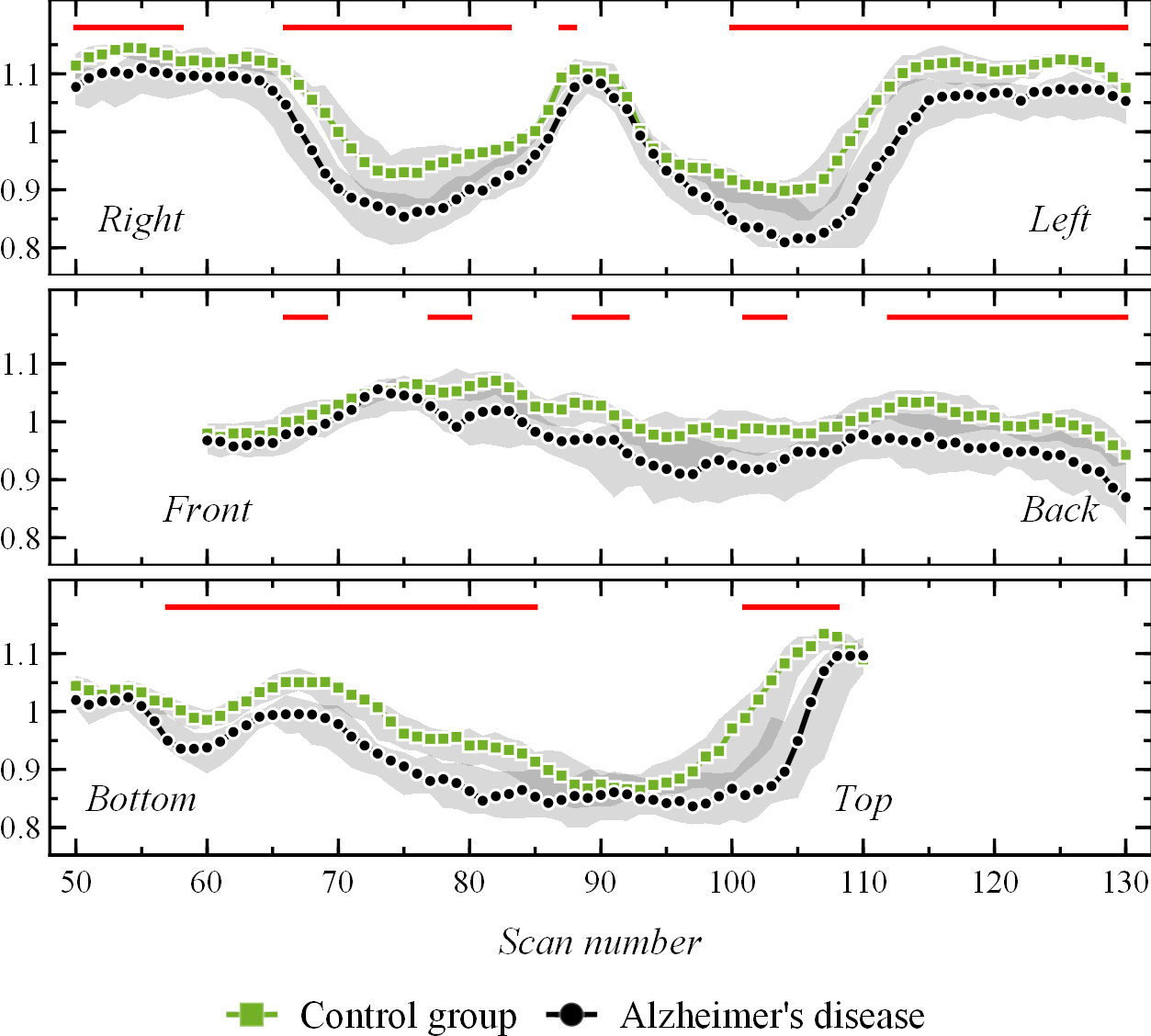}
    \caption{Group-averaged (left) short-scale and (right) long-scale Hurst exponent profiles for the control group (green lines) and Alzheimer's Disease group (black lines) of the ADNI dataset along the x (top), y (middle), and z (bottom) axes. Shaded error bands correspond to 95\% CI of the median. Red horizontal bars indicate significant differences (p<0.01 in Mann-Whitney test). For short-scale exponents, there are no significant differences.}
    \label{fig19}
\end{figure}

Consequently, separately for short and long scales in Figure~\ref{fig19}, we present the Hurst exponent profiles along all three axes, median-averaged over subjects.

It is noteworthy that the median Hurst exponent $H$ assumes values close to or even exceeding one for both short and long scales, indicating a strong persistence of the analyzed signals. This effect is more pronounced at shorter rather than longer scales. On short scales, of particular interest is the disparity between the control and the AD groups: the median Hurst exponent for patients within the AD group typically surpasses that of the control group, although the difference is not statistically significant. This contrast is particularly evident in slices along the $z$-axis (Bottom-Top). At longer scales the situation is reversed: the $H$ estimates for the AD cohort tend to be significantly smaller than for the control group across numerous slices (p<0.01 in two-sided Mann-Whitney test). 
This indicates that in AD local changes in the brain structure influence its scaling properties at larger scales, thereby reorganizing it towards a more variable configuration.

\subsubsection{OASIS and ADNI datasets: Comparison of the Hurst profile for early and mild cognitive impairment patients}

\begin{figure}[H]
    \centering
    \includegraphics[width=0.49\textwidth]{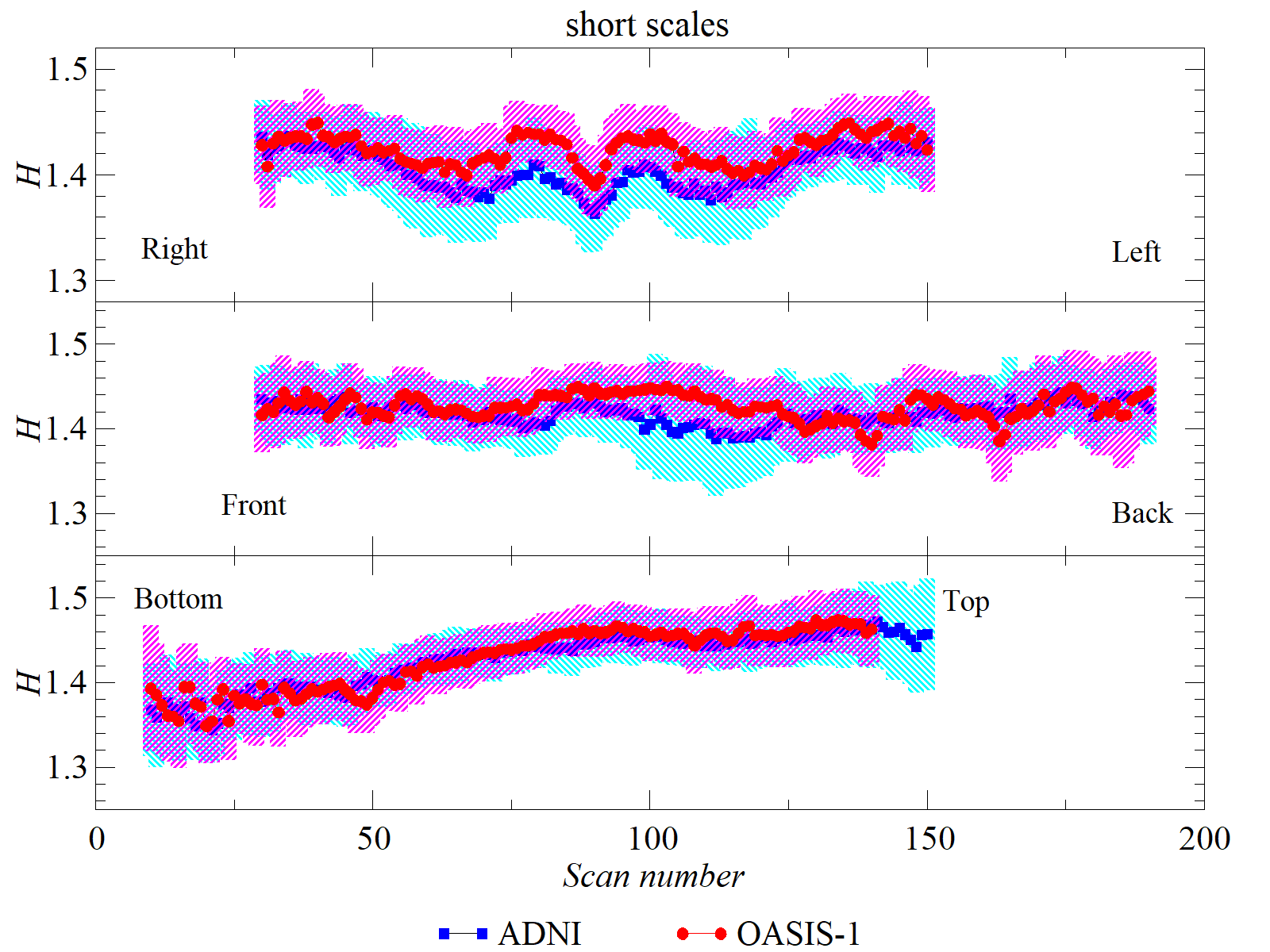}
    \includegraphics[width=0.49\textwidth]{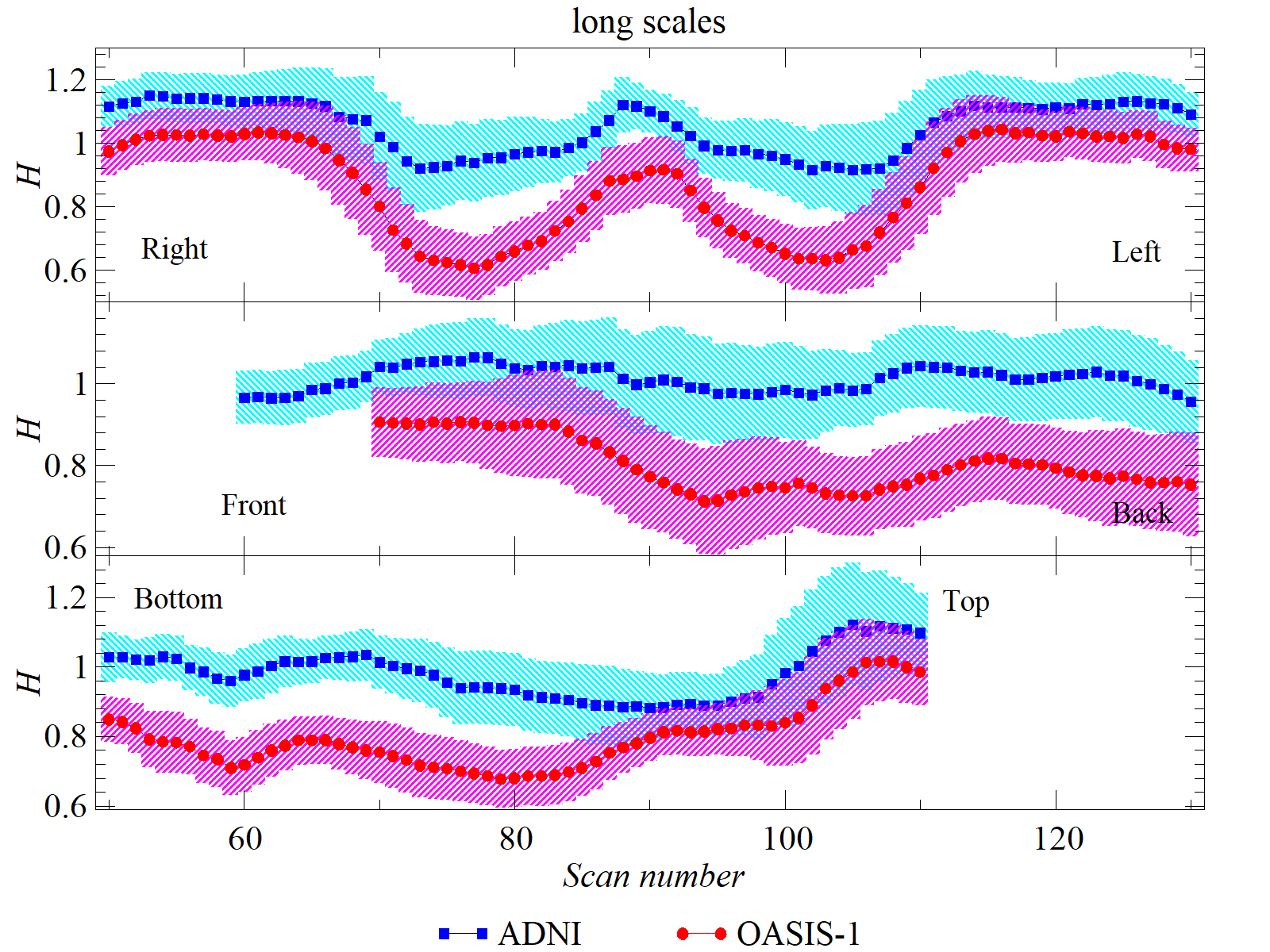}
    \caption{Comparison of ADNI and OASIS-1 datasets between combined group EMCI+MCI (forming a substitute for an early-stage Alzheimer's disease cohort). }
    \label{fig21:FigAlzheimerComparison2}
\end{figure}

We conducted a comprehensive analysis of MRI scans for a combined cohort comprising EMCI and MCI patients sourced from OASIS-1, juxtaposed with analogous data from the ADNI database.
The congruence between the two datasets is visible in Fig.~\ref{fig21:FigAlzheimerComparison2}. The Hurst exponent profiles of each axis and on the two considered scales exhibit similarity across both datasets, thereby affirming the universal scaling properties of the analysis. Discernible disparities in Hurst exponents emerge mainly on long scales, which is potentially attributable to nuances in data preprocessing or variation in MRI scanner parameters. Analogous analyses were conducted for the control groups (see Supplementary Materials), yielding conclusions consistent with those drawn from patient cohorts.

\subsubsection{OASIS dataset: Effects of patient's age on Hurst profiles}

\begin{figure}[H]
    \centering
    \includegraphics[width=0.51\textwidth]{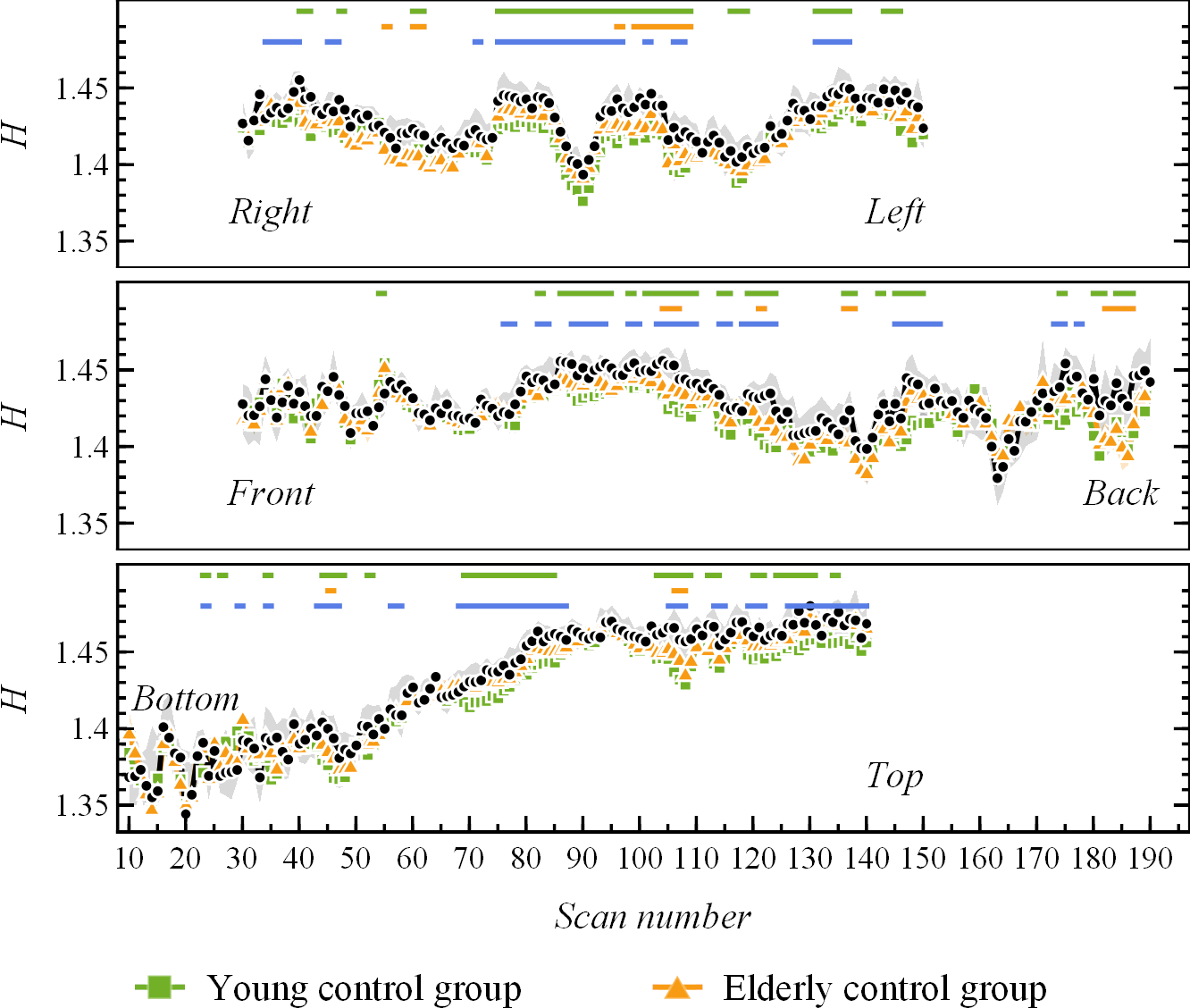}
    \includegraphics[width=0.48\textwidth]{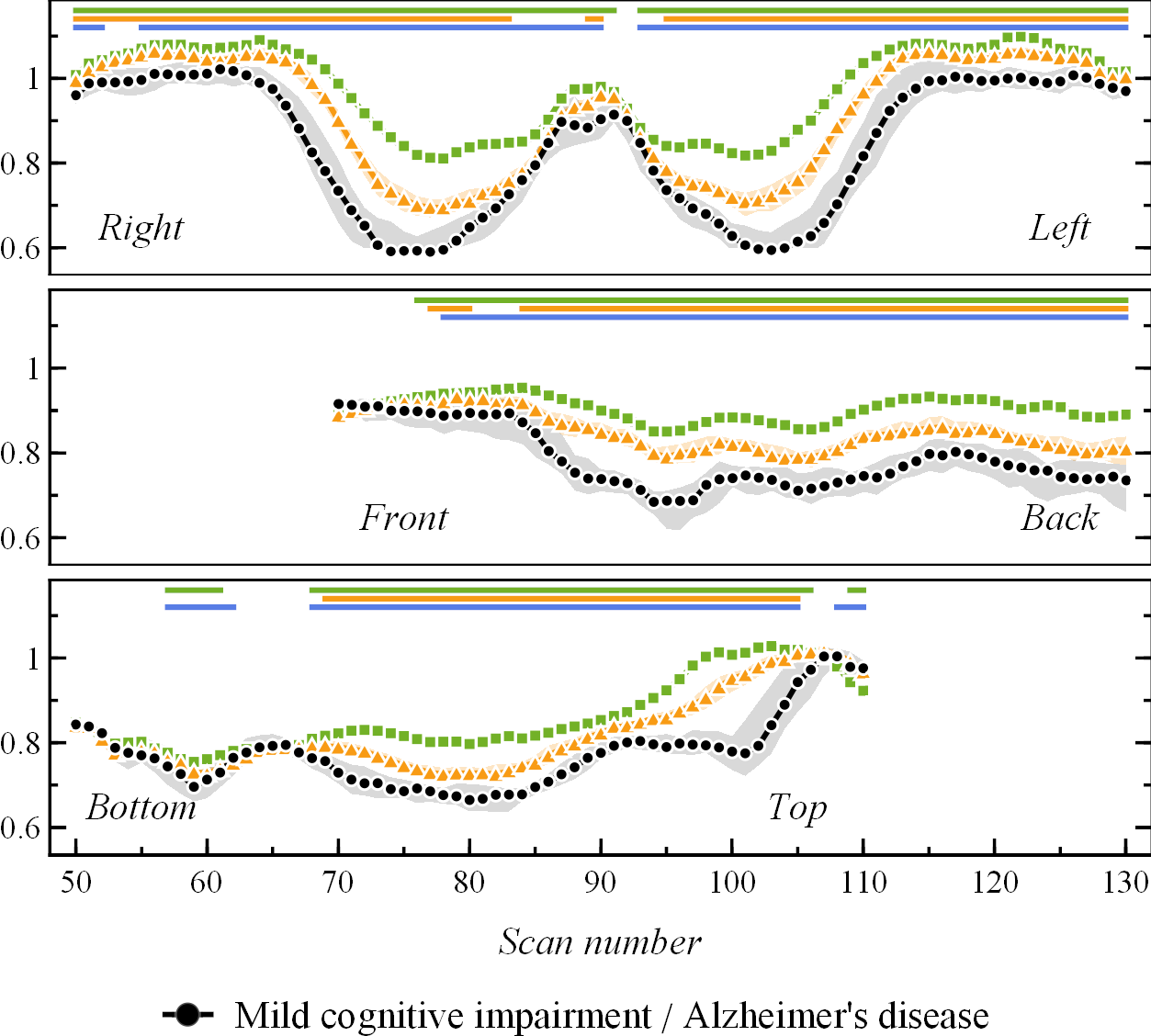}
    \caption{Group-averaged (left) short-scale and (right) long-scale Hurst exponent profiles for the control group (young and elderly) and combined group MCI/AD of the OASIS-1 dataset along the $x$ (top), $y$ (middle) and $z$ (bottom) axes. Shaded error bands correspond to 95\% CI of the median (smaller than markers for the control groups). The horizontal bars indicate pairwise significant differences (p<0.01 in Mann-Whitney test) between yNC and MCI/AD (top, green), eNC and MCI/AD (middle, orange), and yNC and eNC (bottom, blue). Older and cognitively impaired groups have noticeably higher (lower) $H$ values for short-scale (long-scale) exponents, cf. Fig.~\protect\ref{fig19}.}
    \label{fig22:oasisAgeDriftSmall}
\end{figure}

To examine the impact of age on Hurst characteristics, we analyze a cohort of healthy controls divided into two age brackets: young and elderly subjects. Additionally, we contrast these groups with the combined group of patients diagnosed with Mild Cognitive Impairment (MCI) and Alzheimer's Disease (AD). Sizes and mean ages of these groups are given in Tab. \ref{tab:cohort_sizes}. The findings for both short and long scales are depicted in Figure~ \ref{fig22:oasisAgeDriftSmall}.

The between-group difference in the Hurst profiles is more pronounced at the long scales (indeed, the groups significantly differ at $0.01$ level along almost the entire profile). In this scaling regime, the Hurst exponents for young individuals assume larger values than for the elderly subjects along the considered axes ($x$, $y$, and $z$), indicating an age-related trend in the Hurst profiles. 
Moreover, we observe that the trend continues with the MCI/AD group, indicating a relation between low $H$ and cognitive impairment.
On shorter scales, the trend is reversed, as observed in Sec.~\ref{sec:ADNIhurst}, although the examined groups are less distinct (in particular, significant differences between the elderly control and MCI/AD mostly disappear, but there are still extensive regions where the young control statistically significantly differ from both the elderly and MCI/AD).
Specifically, the Hurst exponents for the control group of elderly individuals exhibit values comparable to those of the combined group comprising individuals with MCI and AD.
Moreover, the lowest Hurst values are found among the younger participants within the central brain region.

\subsubsection{OASIS dataset: Effect of early Mild Cognitive Impairment} 
\label{sec:OASIS_MCI_overall}

\begin{figure}[H]
    \centering
    \includegraphics[width=0.49\textwidth]{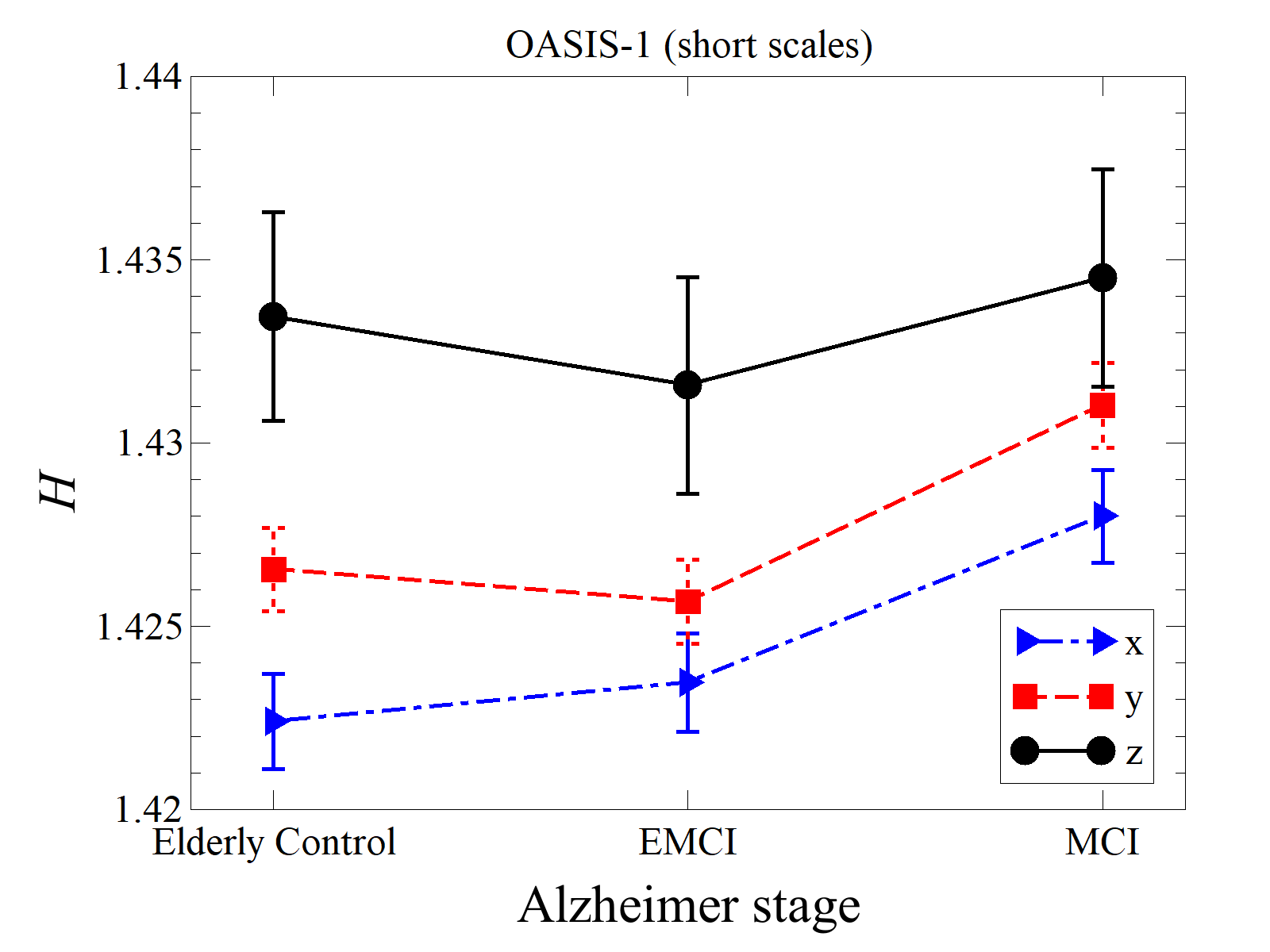}
    \includegraphics[width=0.49\textwidth]{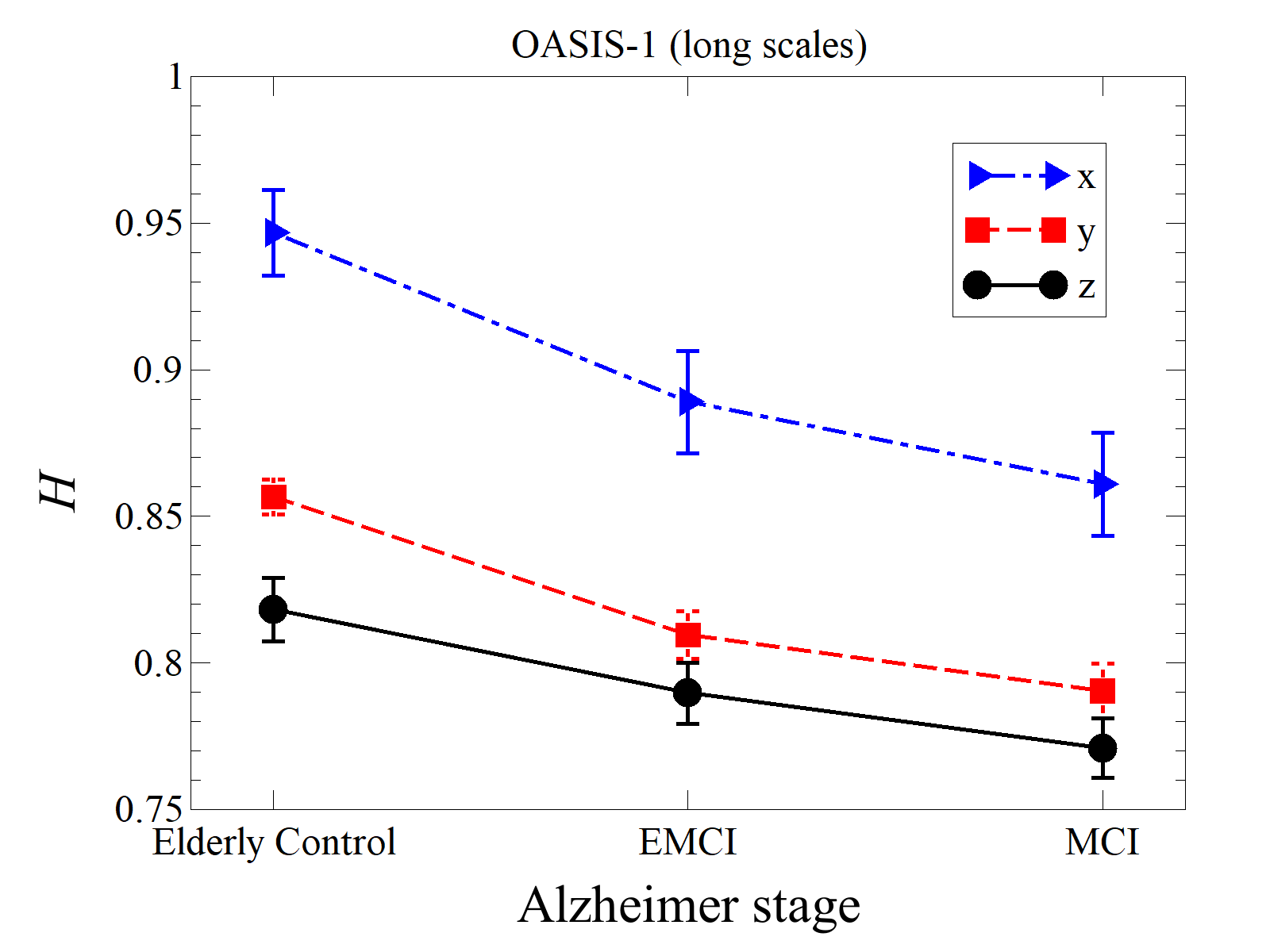}
    \caption{Average Hurst exponents for elderly Control, Early Mild Cognitive Impairment and Mild Cognitive Impairment groups for OASIS-1 dataset (profile medians averaged over cohorts). Error bars denote standard deviations of the medians. For long scales, all differences between eNC and EMCI and between eNC and MCI are statistically significant at $p<0.01$. For short scales, eNC and MCI are significantly different on the \textit{x} ($p<0.05$) and \textit{y} axis ($p<0.1$).}
    \label{fig24:oasisMCIStudy}
\end{figure}

Building on the findings outlined above, we explore the correlation between the Hurst profile and the progression of Alzheimer's disease. Figure~\ref{fig24:oasisMCIStudy} illustrates the average Hurst exponents along the three axes for the elderly control (eNC), EMCI, and MCI groups. Consistently with previous sections, we analyze two scaling regimes. Notably, there is a discernible disparity between the axes, but what captures our attention the most is the volatility of the Hurst exponent within specific directions, which effectively distinguishes between the groups under consideration.
This divergence becomes particularly pronounced for longer scales, where the average Hurst exponent along all axes consistently decreases with the disease progression. Pairwise Mann-Whitney tests reveal significant differences between eNC and EMCI and between eNC and MCI for all axes ($p<0.01$; for individual p-values see Supplementary Table 3). On the short scales, the control and EMCI groups are similar, with the exception of the $x$ axis, where the EMCI group demonstrates a slightly larger Hurst exponent. Additionally, the Hurst exponent noticeably increases for the MCI group. Statistical tests reveal significant differences between eNC and MCI for \textit{x} axis ($p<0.05$) and \textit{y} axis ($p<0.1$; for individual p-values see Supplementary Table 2).

\subsubsection{OASIS dataset: Identification of MCI patients likely to develop dementia}

\begin{figure}[H]
    \centering
    \includegraphics[width=1.0\textwidth]{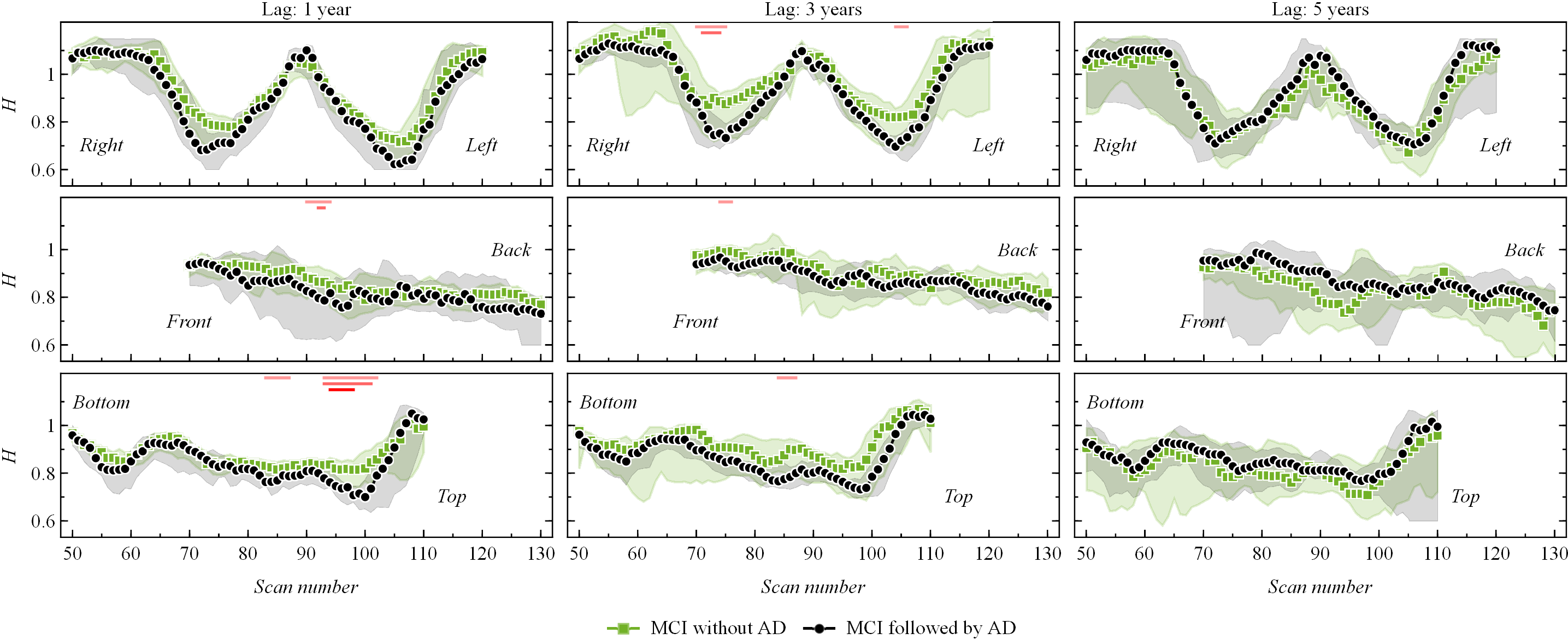}
    \caption{
    Group-average long-scale Hurst exponent profiles for  Mild Cognitive Impairment developing dementia or not within 1,3 or 5 years for OASIS-3 dataset (profile medians averaged over cohorts) along the $x$ (top), $y$ (middle) and $z$ (bottom) axes. Shaded error bands correspond to 95\% CI of the median. The red horizontal bars indicate pairwise significant differences (Mann-Whitney test; p<0.1, 0.05, and 0.01 from top to bottom bar).
    }
    \label{fig25:oasis3MCIStudy}
\end{figure}

We address the clinically relevant task of predicting which patients diagnosed with MCI are likely to develop dementia in the future. We evaluate our method on a subset of the longitudinal OASIS-3 dataset, focusing on patients with mild cognitive impairment who either progress to Alzheimer's disease or do not within a specified time span starting from the date of the scan (see Table~\ref{tab:cohort_sizes} for details). Figure~\ref{fig25:oasis3MCIStudy} illustrates the average Hurst profiles for the long scales along the three axes, highlighting differences between groups progressing to dementia and those who do not within 1, 3, and 5 years. Notably, qualitative similarities can be observed with Figures \ref{fig19}(right) and \ref{fig22:oasisAgeDriftSmall}(right), particularly along the \textit{z}-axis, where the most substantial inter-group differences occur near the dip around slice 100 (p-values $p<0.01$ in Fig.~\ref{fig25:oasis3MCIStudy}) or \textit{x}-axis around slices 70-75 ($p<0.05$).
Importantly, the overall downward drift of the Hurst exponent previously reported in Sec.~\ref{sec:ADNIhurst}-\ref{sec:OASIS_MCI_overall} as characteristic of MCI and dementia, here also serves as a prognostic marker of future development of dementia.
In the regions specified above, the Hurst profiles can differentiate the two groups up to three years prior to the initial diagnosis of dementia (note the sample sizes in Table~\ref{tab:cohort_sizes}).

\subsubsection{Towards FSCA-based diagnostic tools}

In this section, we evaluate the utility of the proposed method as a biomarker for clinically relevant diagnostics. To achieve this, we conduct cross-validation of logistic regression (LR) and support vector machine (SVM) classifiers using FSCA-based input features, comparing them against classifiers based on established biomarkers. Our comparisons include computer-aided diagnostic tools summarized in Tables VI and VII of \cite{MRIreviewBiomarker2018}, with a focus on three binary classification tasks:
\begin{itemize}
    \item Differentiating between healthy controls and patients with dementia (NC vs. AD)
    \item Differentiating between healthy controls and patients with Mild Cognitive Impairment (NC vs. MCI)
    \item Identifying patients diagnosed with Mild Cognitive Impairment who will convert to Alzheimer’s Disease within three years (MCI conv. AD)
\end{itemize}

Table \ref{tab:ml_study} presents a summary of the performance of FSCA-based classification models for these tasks, evaluated on the subsets of ADNI, OASIS-1, and OASIS-3 datasets summarized in Table \ref{tab:cohort_sizes}. For each task, we report the performance of the best model (either LR or SVM).

\begin{table}[h!]
\centering
\begin{tabular}{c|c|ccc}
\hline
task & dataset & accuracy & specificity & sensitivity \\
\hline
NC vs. AD & ADNI & 78\% & 78\% & 78\% \\
MCI vs. AD & ADNI & 72\% & 85\% & 49\% \\
NC vs. MCI & OASIS-1 & 83\% & 90\% & 59\% \\
MCI conv. AD & OASIS-3 & 64\% & 76\% & 49\% \\
\hline

\end{tabular}
 \caption{Performance of FSCA-powered machine learning models for clinically relevant classification tasks: differentiation between the control group and Alzheimer's Disease cohort (NC vs. AD), differentiation between the Mild Cognitive Impairment group and Alzheimer's Disease (MCI vs. AD), differentiation between the control group and Mild Cognitive Impairment cohort (NC vs. MCI), and prediction of conversion from Mild Cognitive Impairment to Alzheimer's Disease within three years (MCI conv. AD).}
    \label{tab:ml_study}
\end{table}

In the \textit{NC vs. MCI} task, our model achieves a mean test accuracy of 83\%, surpassing the average accuracy of $76\% \pm 11\%$ reported in \cite{MRIreviewBiomarker2018}. For the \textit{NC vs. AD} and MCI conversion to AD (\textit{MCI conv. AD}) tasks, the accuracy of FSCA-powered models is 78\% and 64\%, respectively, which fall within the reported average accuracies of $85\% \pm 9\%$ and $73\% \pm 11\%$ in \cite{MRIreviewBiomarker2018}. We report comparable specificity values for the FSCA-based model in the \textit{NC vs. MCI} and \textit{MCI conv. AD} tasks, with average specificity values from \cite{MRIreviewBiomarker2018} being $77\% \pm 15\%$ (NC vs. MCI), $71\% \pm 16\%$ (MCI conv. AD), and $90\% \pm 8\%$ (NC vs. AD). For the \textit{NC vs. AD} task, our model achieves sensitivity values that are consistent with the average sensitivities reported in \cite{MRIreviewBiomarker2018}, which are $77\% \pm 15\%$ (NC vs. MCI), $71\% \pm 16\%$ (MCI conv. AD), and $90\% \pm 8\%$ (NC vs. AD).

These mixed results are anticipated, as both LR and SVM models are not state-of-the-art approaches in computer-aided diagnostic tools \cite{MRIreviewModels2022}. Nonetheless, the presented performance is a promising first step towards the integration of the FSCA method into clinically relevant diagnostic tools, particularly when combined with advanced machine learning models such as gradient boosting or as part of a multimodal approach incorporating biomarkers of different origins.

\subsection{Functional MRI data: fractal properties of brain dynamics}

We conclude the analysis of neuroimaging data by exploring the potential of the FSCA method to capture the intricate temporal dynamics of brain activity in functional magnetic resonance imaging datasets.
This imaging modality is widely recognized as the primary technique for studying brain dynamics, as it tracks changes in the blood oxygen level-dependent (BOLD) signal, serving as a localized, measurable indicator of brain activity. The result is a comprehensive, time-resolved scan that offers a holistic snapshot of brain function. BOLD signals have, however, a non-trivially associated autocorrelation and cross-correlation structure~\cite{ochab2019NPros}
and remain notoriously challenging to analyze due to their very low temporal resolution~\cite{ochab2022}.

In this study, we focus on analyzing fMRI signals obtained from a single healthy human, referred to as the Poldrack dataset~\cite{Poldrack2015}, which comprises sessions involving tasks of varying complexity. Our analysis focuses on the breath-holding task sessions extracted from the revised 1.0.4 version of the dataset. Each session involves approximately 5 minutes of steady breathing, resulting in around 300 snapshots of three-dimensional MRI scans. These sessions are further segmented into behavior-related phases: the inhalation and exhalation phases, and the hold phase, during which the subject holds their breath.

\begin{figure}[H]
    \centering
    \includegraphics[width=0.49\textwidth]{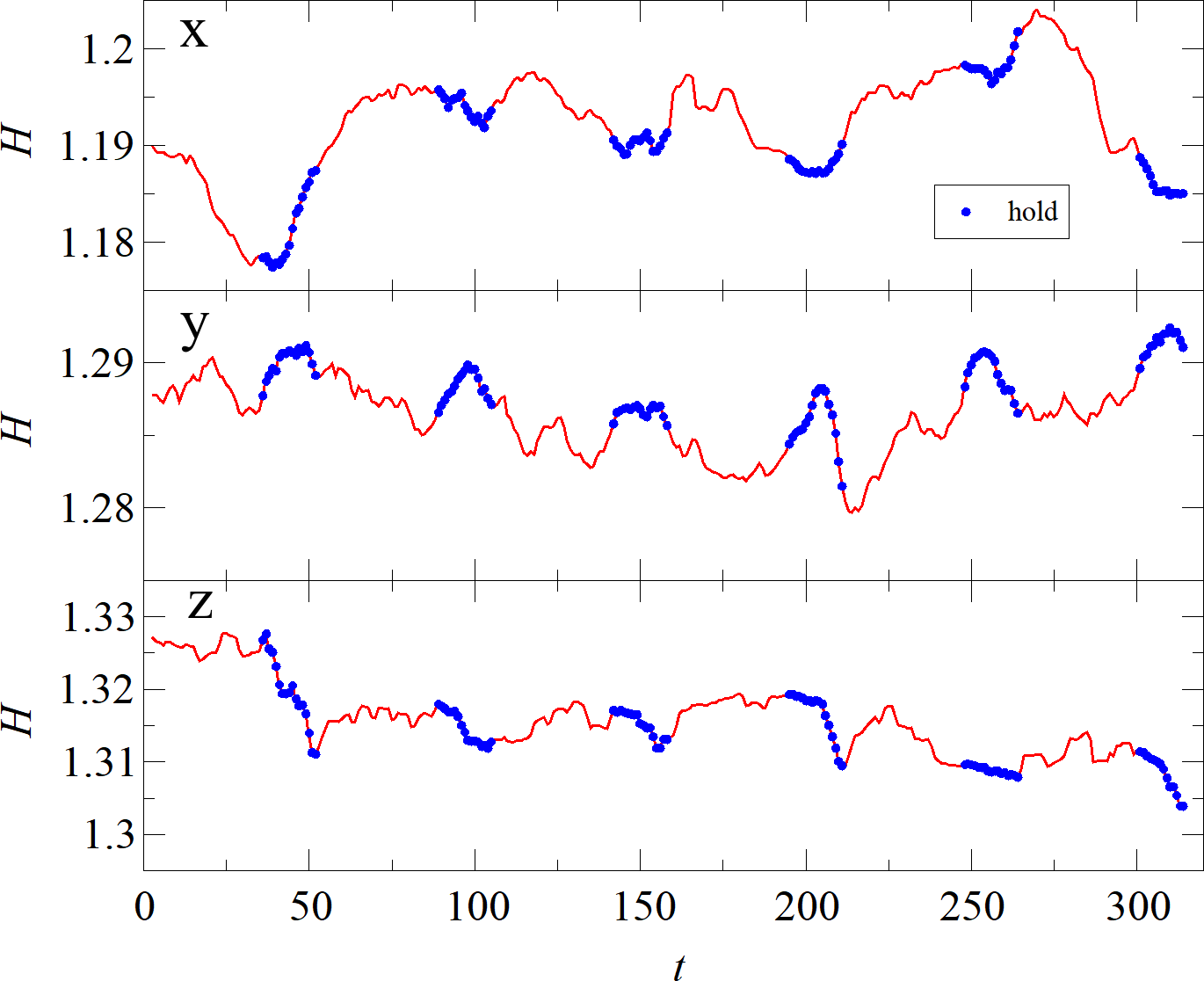}
    \includegraphics[width=0.49\textwidth]{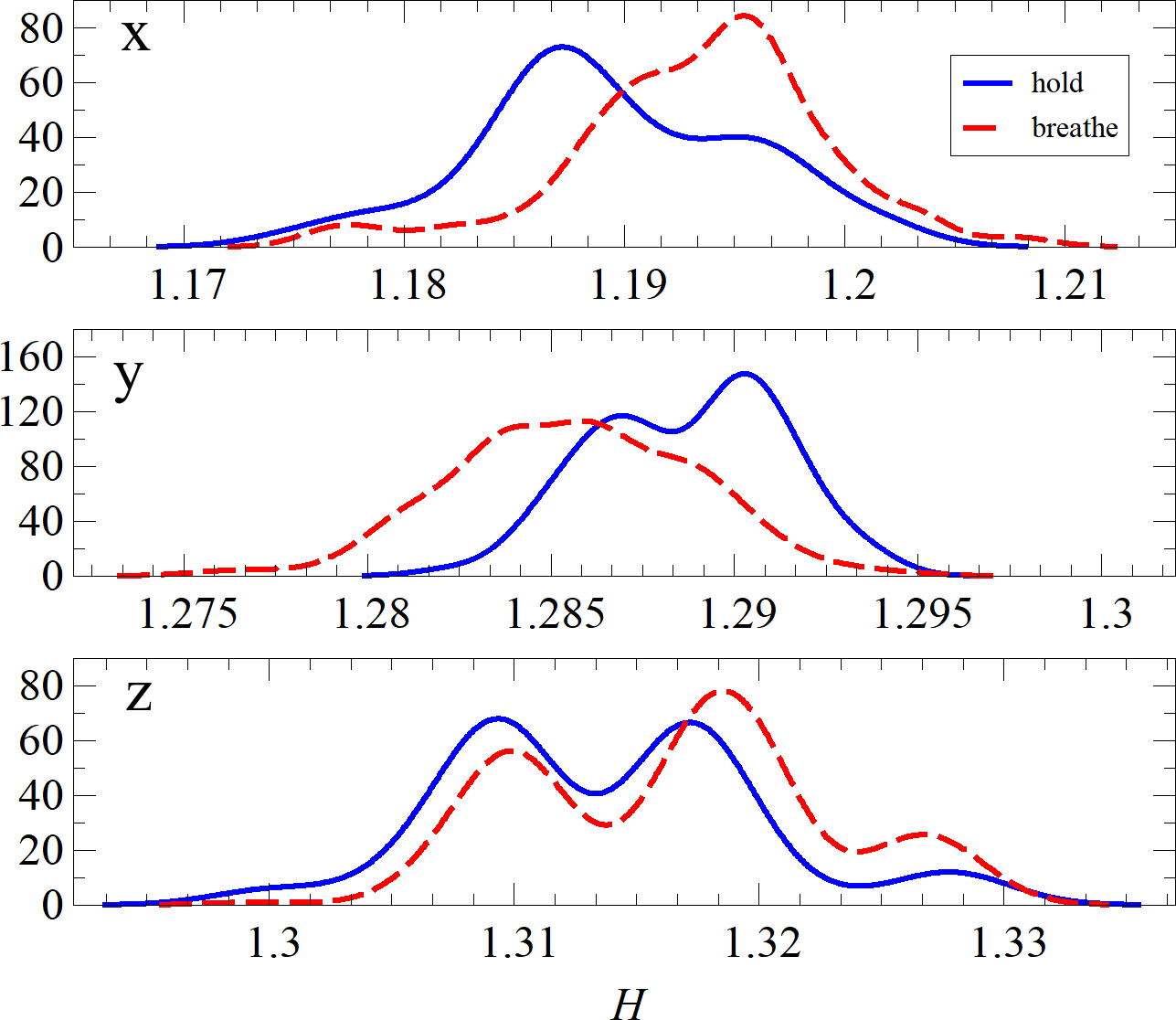}
    \caption{Time variability (left) and histograms (right) of Hurst exponents from a single slice in a single session (Poldrack fMRI data) with long-scale Hurst extraction across three slicing axes (rows). The histograms exhibit a visible breath-dependent shift.} 
    \label{fig25:fig80h800}
\end{figure}

For each snapshot, we decompose the fMRI data into two-dimensional slices along the $x$, $y$, and $z$ axes and apply the FSCA method, yielding a slice-indexed time series of Hurst exponents. In Figure~\ref{fig25:fig80h800}, we illustrate the dynamics of the Hurst exponent for a sample session across the three slicing axes, each for a specific slice. The blue dots represent the Hurst exponent values during the breath-holding phases, forming a discernible pattern within the session.

This pattern is further elucidated in the histograms (cf. Fig~\ref{fig25:fig80h800} right) showing a shift in the distribution of Hurst exponents between breath-taking and breath-holding phases ($p = 2.1 \times 10^{-8}, 2.1 \times 10^{-15}$, and $1.3\times 10^{-6}$, respectively for \textit{x-, y-} and \textit{z}-axis, in two-sided Mann-Whitney test; the effect sizes computed with rank biserial correlation are, respectively, $0.39, -0.55$, and $0.30$). We identify similar alterations in local Hurst exponents across different sessions, though the specific shape of the density plot can vary.




\section{Conclusions}
\label{sec:conclusions}

This paper presents a novel methodology---Fractal Space-Curve Analysis---to analyze geometrically embedded multidimensional data by examining their fractal properties. FSCA first transforms multi-dimensional structures into (one-dimensional) time series via the Hilbert space-filling curve and then employs the detrended fluctuation analysis to quantify its fractal features. We limited this study to 2D SFCs. Below we list the main contributions of the study:
\begin{itemize}
    \item We evaluated the suitability of different types of data linearization: the Hilbert curve, data-driven curves, random data reordering and sweep reordering. In comparison to the Hilbert curve, data-driven SFC proved less sensitive, especially for data exhibiting antipersistence, and significantly more time-consuming, particularly for large datasets. Nevertheless, it may be of some use in the analysis of binary images.
    Random or sweep linearizations retained at best limited information about the underlying fractal properties of the images.
    \item We tested FSCA on synthetic data: 
    two-dimensional fractional Brownian motion (representative of fractal structures), random Cantor sets (exemplifying binary processes), and a dynamical Gaussian process (incorporating temporal dimension).
    The analysis unequivocally demonstrated the method's ability to quantify and discern correlations in two-dimensional images in both stationary and dynamic scenarios. 
    \item We explored different levels of the SFC granularity. Hilbert curve with a low level of nesting (few iterations) was enough to accurately measure the fractal characteristics of the data, allowing even less computation.
    \item We validated the practical utility of the FSCA method for neuroimaging data. We analyzed MRI scans from three databases, ADNI, OASIS-1 and OASIS-3, including patients diagnosed with various levels of dementia and healthy subjects.
    The computational complexity of the FSCA is very favorable compared to state-of-the-art structural MRI tools, taking ${\sim}5$ min. to complete processing an MRI scan. 
    
    \item Hurst exponents on long scales tended to be higher on average in healthy subjects than in patients with Alzheimer's disease. These findings were consistent across all directions of 2D slicing and for all three datasets. Notably, the profile of long-scale Hurst exponents depended on the severity of the disease: decreasing as the disease progresses. The Hurst exponent values appear to correlate also with age and with the likelihood of MCI progressing to dementia. These findings were further supported by an initial work demonstrating promising performance of an FSCA-based diagnostic tool. 
    
    \item To explore the applicability of FSCA to time-dependent neuroimaging data, we applied it to an fMRI dataset from the Poldrack experiment. The method was able to discern shifts in brain activity throughout experimental sessions and, specifically, to distinguish between the breathe-in, breathe-out, and breath-holding phases during the breath-holding task. 
\end{itemize}

In summary, the methodology presented can quantitatively describe the spatial and temporal organization of neuroimaging data, opening opportunities for both research and clinical applications. Furthermore, very few assumptions on FSCA allow it to be utilized in other scientific fields where the characterization of multidimensional data is required.

\section*{Acknowledgements}

This work was carried out within the research project ‘‘Bio-inspired artificial neural networks’’
(grant no. POIR.04.04.00-00-14DE/18-00) within the Team-Net
program of the Foundation for Polish Science co-financed by the
European Union under the European Regional Development Fund.
The research for this publication has been supported by a grant from the Priority Research Area DigiWorld under the Strategic Programme Excellence Initiative at Jagiellonian University.
JJ acknowledges the support of the Faculty of Physics, Astronomy, and Applied Computer Science of the Jagiellonian University through grant no. LM/3/JJ.
IC acknowledges the support of MICINN (Spain) grant no. PID2021-125534OB-I00.

Data were provided [in part] by OASIS-1: Cross-Sectional: Principal Investigators: D. Marcus, R, Buckner, J, Csernansky J. Morris; P50 AG05681, P01 AG03991, P01 AG026276, R01 AG021910, P20 MH071616, U24 RR021382.

Data collection and sharing for the Alzheimer's Disease Neuroimaging Initiative (ADNI) is funded by the National Institute on Aging (National Institutes of Health Grant U19 AG024904). The grantee organization is the Northern California Institute for Research and Education. In the past, ADNI has also received funding from the National Institute of Biomedical Imaging and Bioengineering,  the Canadian Institutes of Health Research, and private sector contributions through the Foundation for the National Institutes of Health (FNIH) including generous contributions from the following: AbbVie, Alzheimer’s Association; Alzheimer’s Drug Discovery Foundation; Araclon Biotech; BioClinica, Inc.; Biogen; Bristol-Myers Squibb Company; CereSpir, Inc.; Cogstate; Eisai Inc.; Elan Pharmaceuticals, Inc.; Eli Lilly and Company; EuroImmun; F. Hoffmann-La Roche Ltd and its affiliated company Genentech, Inc.; Fujirebio; GE Healthcare; IXICO Ltd.; Janssen Alzheimer Immunotherapy Research \& Development, LLC.; Johnson \& Johnson Pharmaceutical Research \& Development LLC.; Lumosity; Lundbeck; Merck \& Co., Inc.; Meso Scale Diagnostics, LLC.; NeuroRx Research; Neurotrack Technologies; Novartis Pharmaceuticals Corporation; Pfizer Inc.; Piramal Imaging; Servier; Takeda Pharmaceutical Company; and Transition Therapeutics.



\section*{Data availability statement}
Data are available from the original sources:~\cite{ADNI,OASIS1,Poldrack2015}.
Code is available in \url{https://github.com/Mark-Kac-Center/FSCA}.
Correspondence should be addressed to Jacek Grela (jacek.grela@uj.edu.pl).

\section*{Conflict of interests}
The authors declare no competing interests.


% ...

\bibliography{References} 
\bibliographystyle{iopart-num} 

\appendix
\renewcommand{\figurename}{Supp. Figure}
\renewcommand\thefigure{\arabic{figure}}
\renewcommand{\tablename}{Supp. Table}
\renewcommand\thetable{\arabic{table}}
\setcounter{figure}{0}    
\setcounter{table}{0}    

\section*{Supplementary information}

\begin{figure}[H]
    \centering
    \includegraphics[width=1\textwidth]{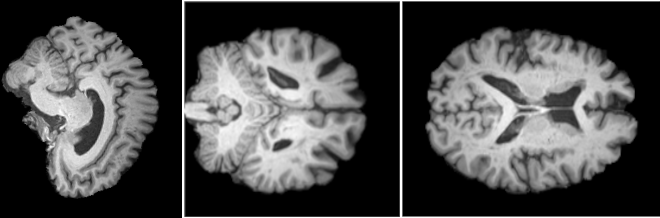}
    \caption{A sample MRI scan (ADNI dataset) along the $x$ (left-right or frontal), $y$ (front-back or saggital) and $z$ (bottom-top or longitudinal) axis.}
    \label{fig18:FigMRIScan}
\end{figure}

\begin{figure}[H]
    \centering
    \includegraphics[width=\columnwidth]{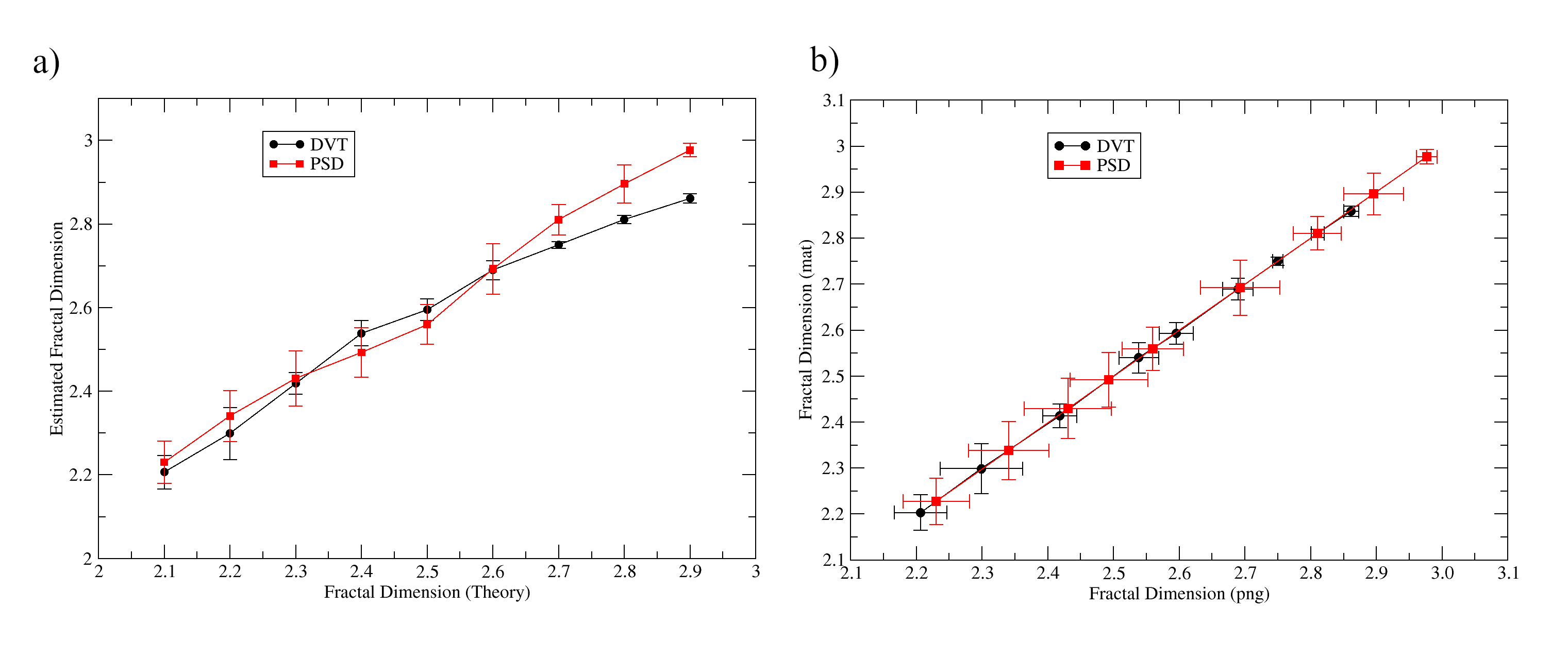}
    \caption{Results of analysis of fractional Brownian motions. a)  Comparison between fractal dimension $D$ estimated by means of DVT and PSD algorithms. b) Comparision $D$ estimated for data in different computer formats and by means of different algorithms: data matrix (mat) and picture (png).}
    \label{sup_fig01:Brownian_comparison}
\end{figure}

\begin{figure}[H]
    \centering
    \includegraphics[width=0.7\textwidth]{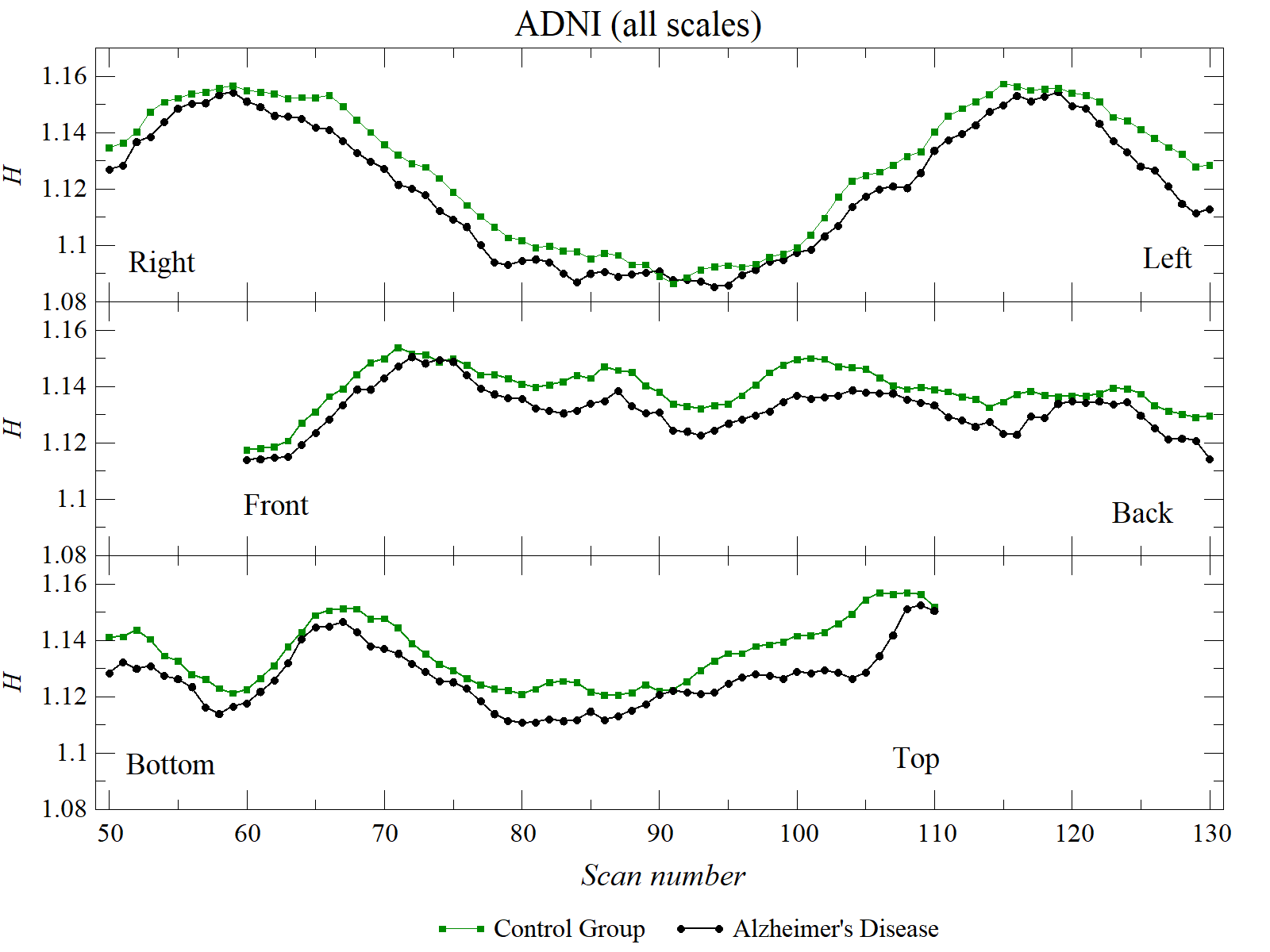}
    \caption{Median of the Hurst exponent (all scales) of the Hilbert curve estimated along the $x$, $y$, $z$-axis (ADNI).}
    \label{sup_fig02}
\end{figure}

\begin{figure}[H]
    \centering
    \includegraphics[width=0.49\textwidth]{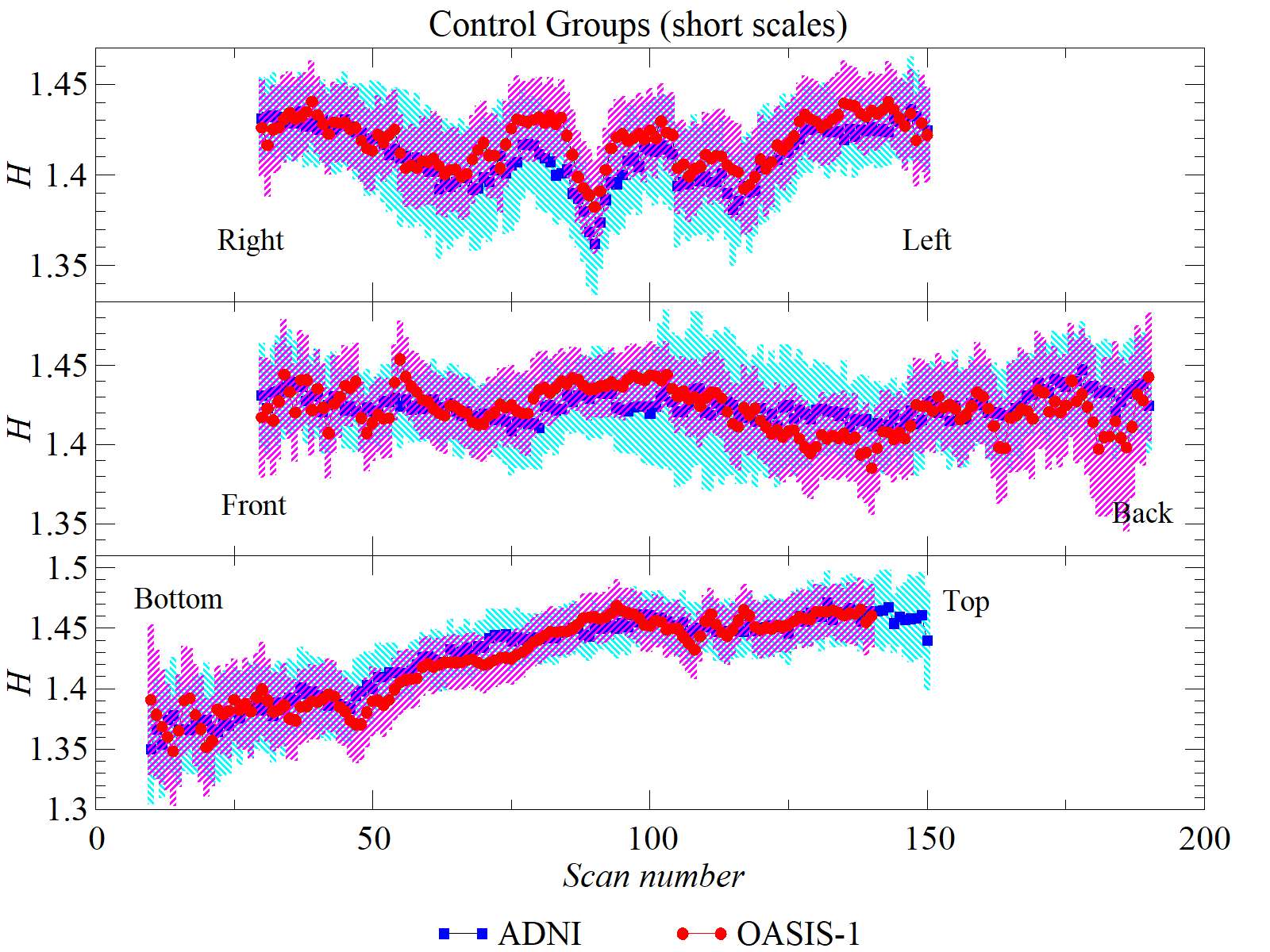}
    \includegraphics[width=0.49\textwidth]{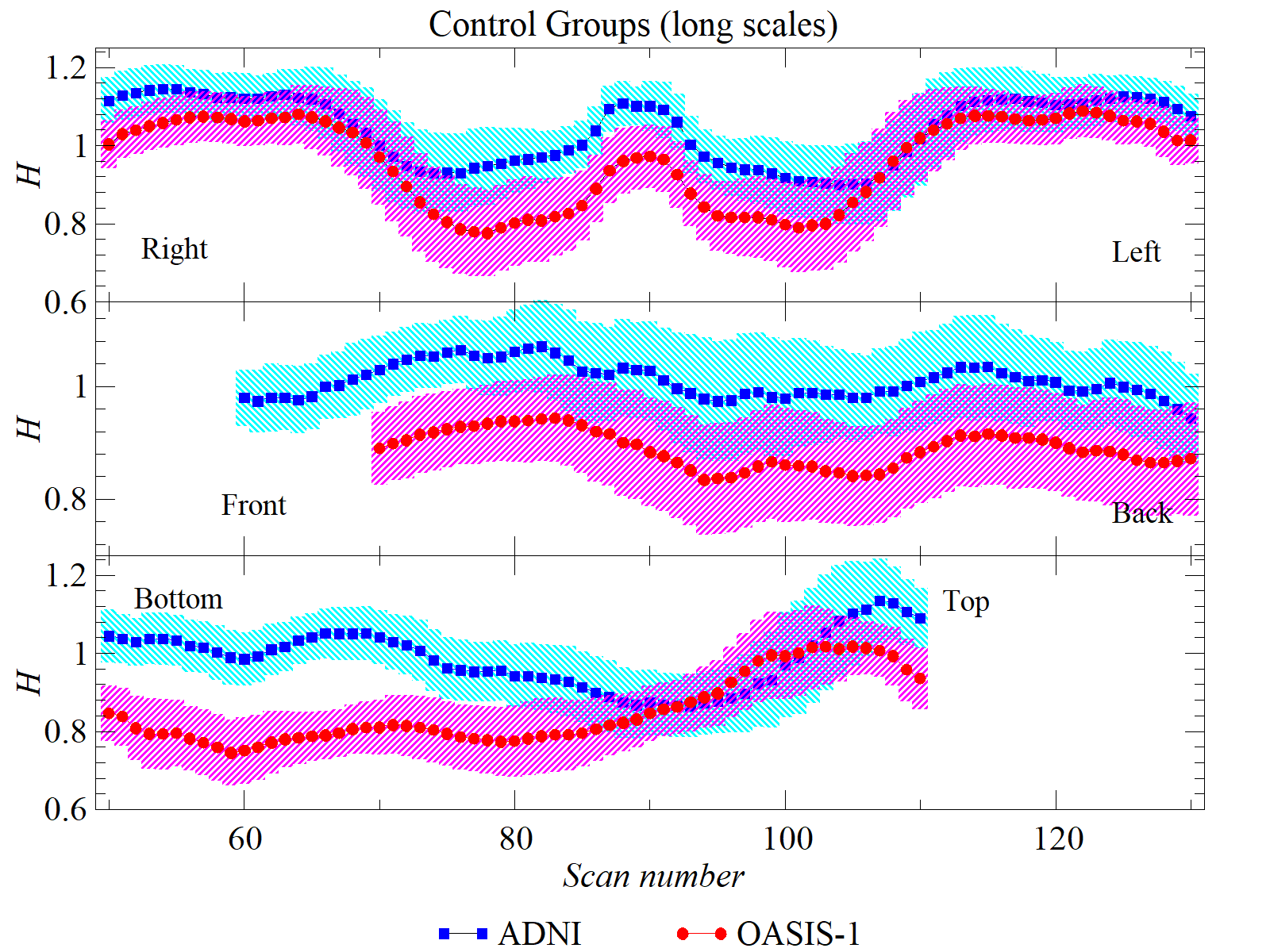}
    \caption{Comparison of control groups in the ADNI and OASIS-1 datasets.}
    \label{sup_fig03:FigAlzheimerComparison}
\end{figure}

:

\begin{table}[h!]
\begin{tabular}{|l|l|l|l|l|l|}
\hline
\multicolumn{6}{|c|}{\textbf{artificial 2D data}} \\
\hline
\textbf{data dim} & 16x16 & 32x32 & 64x64 & 128x128 & 256x256\\
\textbf{exec time} [s] & 0.01657(52) & 0.0206(12) & 0.0367(11) & 0.0854(15) & 0.2600(33)  \\
\hline
\textbf{data dim} & 512x512 & 1024x1024 & 2048x2048 & & \\
\textbf{exec time} [s] & 0.904(72) & 4.03(30) & 16.31(19) & &\\
\hline
\hline
\multicolumn{6}{|c|}{\textbf{artificial 3D data}} \\
\hline
\textbf{data dim} & 8x8x8 & 16x16x16 & 32x32x32 & 64x64x64 & 128x128x128 \\
\textbf{exec time} [s] & 0.03178(13) & 0.100(13) & 0.6355(80) & 4.643(22) & 38.18(18) \\
\hline

\end{tabular}
    \caption{FSCA time performance (using the Hilbert SFC) on artificial two- and three-dimensional data.}
    \label{tab:my_label}
\end{table}

\begin{table}[]
\begin{tabular}{l|cc|cc|cc}
                          & \multicolumn{2}{c|}{x} & \multicolumn{2}{c|}{y} & \multicolumn{2}{c}{z} \\
                          & EMCI     & MCI         & EMCI      & MCI        & EMCI      & MCI       \\ \hline
\multicolumn{1}{r|}{EC}   & 0.21     & 0.015**     & 0.43      & 0.031*     & 0.85      & 0.27      \\
\multicolumn{1}{r|}{EMCI} & -        & 0.10        & -         & 0.23       & -         & 0.48     
\end{tabular}
\caption{Statistical significance of pairwise differences in OASIS-1 dataset for short-scale Hurst exponents, see Fig. 25 (right). P-values from the Mann-Whitney test are uncorrected for multiple comparisons. Significance codes: \textbf{*} $p<0.1$, \textbf{**} $p<0.05$ after Bonferroni correction.}
\label{tab:sig_short}
\end{table}

\begin{table}[h!]
\begin{tabular}{l|cc|cc|cc}
                          & \multicolumn{2}{c|}{x}                                  & \multicolumn{2}{c|}{y}                                  & \multicolumn{2}{c}{z}                                 \\
                          & \multicolumn{1}{c}{EMCI}   & \multicolumn{1}{c|}{MCI}   & \multicolumn{1}{c}{EMCI}   & \multicolumn{1}{c|}{MCI}   & \multicolumn{1}{c}{EMCI}   & \multicolumn{1}{c}{MCI}  \\ \hline
\multicolumn{1}{r|}{EC}   & \footnotesize$2.8\times10^{-6}$*** & \footnotesize$1.8\times10^{-5}$*** & \footnotesize$1.5\times10^{-6}$*** & \footnotesize$3.1\times10^{-6}$*** & \footnotesize$8.6\times10^{-4}$*** & \footnotesize$1\times10^{-4}$*** \\
\multicolumn{1}{r|}{EMCI} & -                          & \footnotesize0.28                       & -                          & \footnotesize0.10                       & -                          & \footnotesize0.13                    
\end{tabular}
\caption{Statistical significance of pairwise differences in OASIS-1 dataset for long-scale Hurst exponents, see Fig. 25 (right). P-values from the Mann-Whitney test are uncorrected for multiple comparisons.  Significance codes: \textbf{***} $p<0.01$ after Bonferroni correction.}
\label{tab:sig_long}
\end{table}

\end{document}